\documentclass[twocolumn, english,showpacs,preprintnumbers,amsmath,amssymb]{revtex4}

\usepackage{amsmath}
\usepackage{amssymb}
\usepackage{graphicx}
\usepackage{babel}
\usepackage{times}

\allowdisplaybreaks[3]

\begin{document}

\title{ Renyi entropies for classical stringnet models}

\author{M. Hermanns}

\author{S. Trebst}

\affiliation{Institute for Theoretical Physics of Cologne, 50937 Cologne,
Germany}

\date{\today}

\begin{abstract}
In quantum mechanics, stringnet condensed states -- a family of prototypical states exhibiting non-trivial topological order --
can be classified via their long-range entanglement properties, in particular topological corrections to the prevalent area law of the entanglement entropy.
Here we consider classical analogs of such stringnet models whose partition function is given by an equal-weight superposition of classical stringnet configurations. Our analysis of the Shannon and Renyi entropies for a bipartition of a given system reveals that the prevalent volume law for these classical entropies is augmented by subleading topological corrections that are intimately linked to the anyonic theories underlying the construction of the classical models. We determine the universal values of these topological corrections for a number of underlying anyonic theories including su(2)$_k$, su(N)$_1$, and su(N)$_2$ theories.

\end{abstract}
\pacs{65.40.gd, 89.70.Cf, 05.20.-y}
\maketitle


\section{Introduction}
\label{Introduction}
The concept of topological order  -- long-range order beyond the conventional paradigm of symmetry broken order -- has profoundly broadened our view of strongly correlated systems over the last decades \cite{toporder}. 
Archetypal realizations of actual materials exhibiting such unconventional order include the (fractional) quantum Hall liquids \cite{tsui} discovered over 30 years ago, Sr$_2$RuO$_4$ as a possible realization of a $p_x + i p_y$ superconductor \cite{PSC} or in the more recent past the discovery of topological band insulators \cite{TopoInsulators}. Further insight into the rich physics of topologically ordered systems has been achieved by the analytical understanding of exactly solvable spin models including the toric code \cite{ToricCode} or Kitaev's honeycomb model \cite{HoneycombModel}.
 
\begin{figure}[b]
  \begin{centering}
  \includegraphics[width=.6\columnwidth]{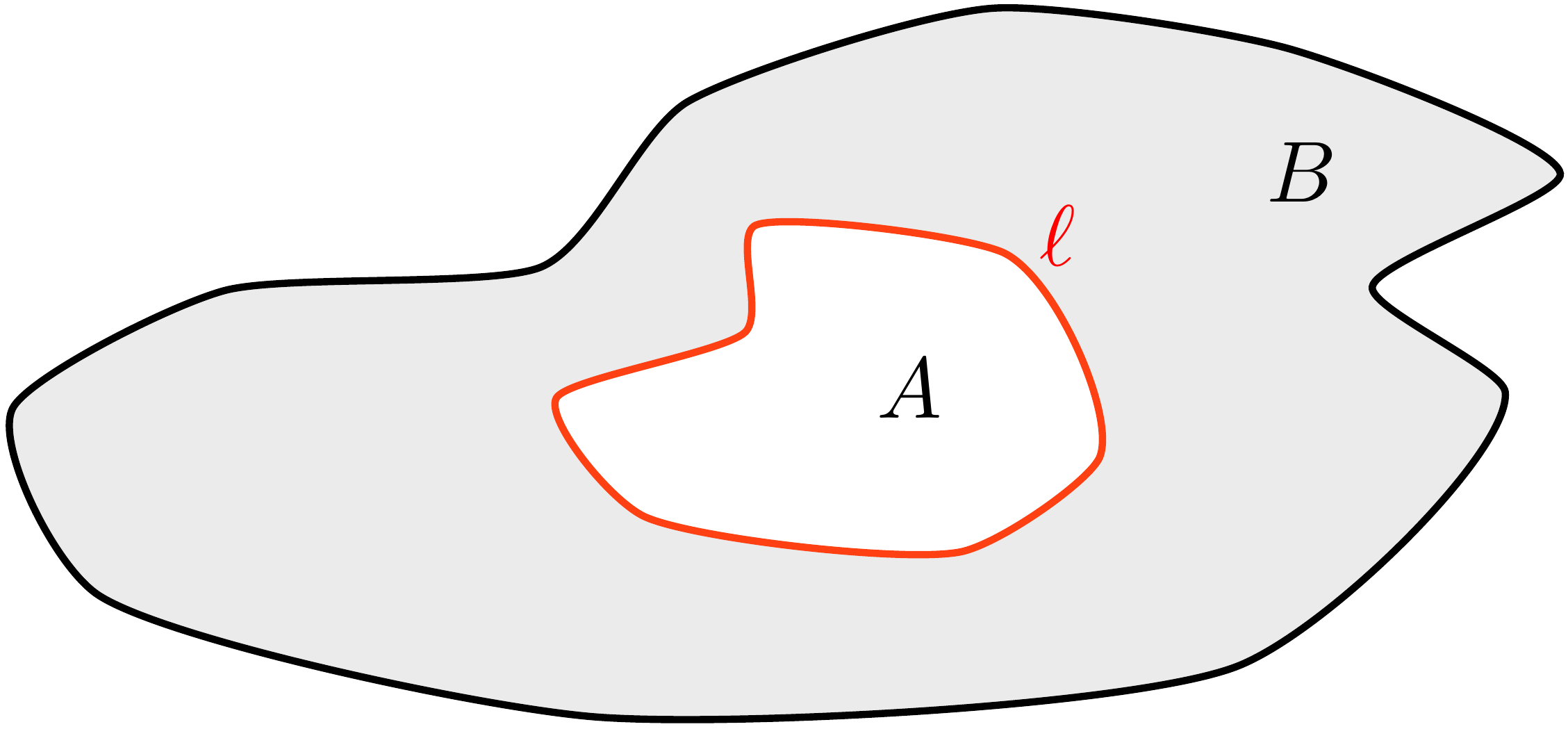}
  \end{centering}
  \caption{(color online) Bipartition of a quantum system.}
  \label{Fig:Partition}
\end{figure}

Despite all this progress it has remained a challenging task to unambiguously identify a topologically ordered state in a given system -- both in experiments or theoretical approaches. On the analytical side, concepts from quantum information theory have proven particularly helpful in generating a powerful measure of topological order in model systems. A key insight is that the topological order present in a quantum many-body system is reflected in its entanglement properties. In its most pronounced incarnation the long-range order present in topologically ordered systems leads to long-range entangled states \cite{LongRangeEntanglement}, a scenario that applies to most of the examples of topologically ordered states above. The notable exception are the topological band insulators which exhibit short-range entanglement \cite{ShortRangeEntanglement}. Here we will focus on the former class of long-range entangled states.
A key concept to measure entanglement in a quantum many-body system is the entanglement entropy calculated by dividing the system into two parts $A$ and $B$ as illustrated in Fig.~\ref{Fig:Partition}, obtaining the reduced density matrix of one partition by tracing out the other, e.g. $\hat\rho_A=\mbox{Tr}_B(|\psi\rangle\langle\psi|)$, and then collapsing the information in the reduced density matrix into the so-called entanglement entropy 
\[
     S_{A}=-\mbox{Tr}_A[\hat\rho_A\log \hat\rho_A] \,.
\]

Despite the relatively broad-brush character of the entanglement entropy studying its dependence on the geometry of the bipartition has been shown to allow for a rather general classification of ground states of interacting quantum many-body systems. For systems restricted to two spatial dimensions the entanglement entropy exhibits a so-called `area law' or more specifically a `boundary law' for all gapped ground states; it grows with the length $\ell$ of the boundary between the partitions $A$ and $B$
\begin{align}
\label{eq:entropy quantum}
S_{A} & =c_\ell \ell-\gamma_{\rm quantum}+\ldots
\end{align}
where any additional terms indicated by the dots are subleading terms of order $O(1/\ell)$, i.e. vanish in the limit $\ell\rightarrow \infty$. 
One of the more striking features of this boundary law is the occurrence of a subleading constant contribution that for a smooth boundary, i.e. one without any sharp corners, is independent of the size or the geometry of the bipartition and thus indicates long-range entanglement.  
This constant  $\gamma_{\rm quantum}$ is often called the {\em topological entropy} \cite{KitaevPreskill,LevinWenEntropy}, since it strictly vanishes for disordered or conventionally ordered states while for a topologically ordered state it always remains finite. 
In fact, the topological entropy has to lock into a universal value, which is tightly connected to the effective topological field theory describing the non-local topological order in the system at hand. It was shown \cite{KitaevPreskill,LevinWenEntropy} that
 \begin{align}
 \gamma_{\rm quantum}= \log \mathcal{D},
 \label{eq:top entropy quantum}
 \end{align} where $\mathcal{D}$ is the so-called total quantum dimension, an important characteristic of the effective topological field theory. 
An alternative perspective to understand the origin of a finite constant $\gamma_{\rm quantum}$ for a topologically ordered state emerges from considering the entropic contribution arising solely from the boundary. In contrast to a conventionally ordered  or  entirely disordered system, the states on the boundary of a topologically ordered system are found to be subtly  constrained which then leads to a non-trivial entropic contribution of this boundary. 
We will discuss details of both perspectives in the remainder of the manuscript.

Despite the obvious fact that classical systems do not exhibit any entanglement, the notion of the entanglement entropy has a straight-forward companion for classical statistical systems -- the Shannon entropy, which we can analogously define for a bipartition of the system as illustrated in Fig.~\ref{Fig:Partition}. It is then given as
\begin{align}
S_{A} & = -\sum_{\{j_{A}\}}p_{j_{A}}\log p_{j_{A}} \, ,
\label{eq:Shannon intro}
\end{align}
where  $\{j_{A}\}$ denotes the set of possible classical configurations in $A$ and $p_{ j_{A} }$ is the statistical probability of a given configuration $j_{A}$. 

The analogy to the entanglement entropy goes even one step further, as the Shannon entropy also exhibits a rather characteristic scaling with the size and the geometry of the bipartition. In its most general form it follows a `volume law' of the form
\begin{align}
S_{A} & =c_{V}V_{A}+c_{\ell}\ell-\gamma_{\rm classical}+\ldots\, ,
\end{align}
 where $V_{A}$ is the volume of partition $A$ and $\ell$ is again the length of the boundary of the partition. The dots indicate subleading terms that vanish in the limit of $V_A, \ell\rightarrow \infty$. 
 
The occurrence of a non-vanishing constant term $\gamma_{\rm classical}$ in this volume law for the classical Shannon entropy is again tightly connected to the occurrence of long-range information in the classical system. A prototypical family of such classical systems exhibiting long-range information are classical variants of so-called stringnet states, which we will introduce and discuss in much detail in the following. The main result of this manuscript is that for this broad variety of classical systems we establish a universal equation relating the constant term $\gamma_{\rm classical}$ to a characteristic feature of the topological field theory underlying these stringnet states. The methods, which allow us to derive this relation, are based on recent work by Fendley and Simon \cite{SimonFendley}. 

We should note that  a related question was discussed previously by Castelnovo and Chamon in \cite{CastelnovoChamon1}, where the authors studied  classical variants of a class of quantum double models, including  Kitaev's toric code \cite{ToricCode}. The authors found that the values of the classical and topological entropy are identical for this class of models. This statement is, in general, no longer true for the classical variants of the stringnet states.  Instead, we find that
\begin{align}\label{eq:top term}
\gamma_{\rm classical}&=\log M \, ,
\end{align}
where $M$ is the number of Abelian particles in the anyonic theory underlying the stringnet states.  Our results shed new light on the question on what kind of information in the quantum model can be retained in the corresponding classical variant. 

Our discussion of the above results is structured as follows in the remainder of the manuscript:
We provide a general introduction to the classical Shannon and Renyi entropies in Section~\ref{sec:Shannon entropy}. 
In Section~\ref{sec:classical_stringnet_models} we introduce the stringnet models, first in the context of quantum double models based on the so-called su(2)$_k$ anyon theories, and then describe their classical analogs.
In Section~\ref{sec:crossing_symmetry} we outline the methods we use to analytically compute the Renyi entropies for these classical stringnet analogs with the results for the Renyi entropies and the topological entropy of Eq.~\eqref{eq:top term} presented in Section~\ref{sec:Renyi entropy}. The manuscript finishes with an outlook on more general stringnets, a summary of our main results, and several appendices providing the details of our calculation.


\section{Classical Shannon and Renyi entropy}
\label{sec:Shannon entropy}

In quantum mechanics, one of the most important measures of entanglement in a system of multiple quantum mechanical degrees of freedom is the entanglement entropy. We will briefly recap this notion in the following and then turn to the classical analogs of these concepts.

\subsection{Entanglement entropy}

Let us consider a quantum system with ground state $|\psi\rangle$ and a bipartition along a smooth cut (without any sharp corners) that divides the system into two parts called $A$ and $B$. 
As already outlined in the introduction we can characterize the entanglement of the two partitions by calculating the reduced density matrix of one of the two partitions, say partition $A$, by tracing out the other one to obtain
\[
	\hat\rho_A=\mbox{Tr}_B(|\psi\rangle\langle\psi|) \,.
\]
We can now readily diagonalize this reduced density matrix to obtain its eigenvalues $p_j$, which are the probabilities to find system $A$ in the quantum state corresponding to the respective eigenvector. In terms of these eigenvalues, the entanglement entropy can thus be readily calculated as 
\begin{align}
S_{A} =-\sum_{j}p_{j}\log p_{j} \,.
\label{eq:entanglement entropy}
\end{align}
Note that for a quantum system the so-calculated entanglement entropies $S_A$ and $S_B$ are equal, i.e. $S_A = S_B$.

\subsection{Classical entropies}

To identify a classical analog to the entanglement entropy we note that the definition of the entanglement entropy in the form of Eq.~\eqref{eq:entanglement entropy} has a straightforward interpretation also for classical systems. In particular, we can replace the quantum mechanical probabilities $p_j$ to find the system in a certain state $j$ by their classical counterparts to find subsystem $A$ in a particular classical configuration $j_A$. The classical analog of the entanglement entropy is then simply given by the well-known Shannon entropy
\begin{align}
   S_{A} & = -\sum_{\{j_{A}\}}p_{j_{A}}\log p_{j_{A}} \, ,
\label{eq:partition entropy}
\end{align}
where the sum now runs over the set of all possible configurations $\{j_{A}\}$ of subsystem $A$.
Similar to the quantum mechanical case, the probability of a given classical configuration $j_A$ is obtained 
by summing over all possible complementary configurations in partition $B$, i.e.
\begin{align}
p_{\{j_{A}\}} & =\frac 1 Z \sum_{\{j_{B}\}}e^{-E(\{j_{A+B}\})/(k_B T)} \,, \label{eq:classical probability}
\end{align}
where $E(\{j_{A+B}\})$ is the energy of the configuration in the full system, $k_B$ is the Boltzmann factor,
and $T$ the temperature. 
Note that in general the classical Shannon entropy is not equivalent for the two partitions $A$ and $B$, i.e. $S_A \neq S_B$.

Instead of directly calculating the Shannon entropy \eqref{eq:partition entropy} it is often more convenient to compute 
one of the Renyi entropies 
\begin{align}
\label{eq:Renyi_entropy_def}
S^{(n)}_A=\frac 1 {1-n}\log\left[\sum_{\{j_{A}\}} p_{j_A}^n \right] \,,
\end{align}
where the index $n$ typically is an integer $n \geq 2$.
The Renyi entropies are bounded by each other by $S_{n'}\leq S_{n}$ for $n'>n$, and recover the Shannon entropy in the 
limit $n\rightarrow 1$. 
They can be computed by considering $n$ copies (replicas) of the system \cite{ReplicaTrick}. In doing so, the different replicas of part $A$ are required to have identical configurations, while the configurations in the replicas of part $B$ are independent of one another, 
see e.g. Ref.~\cite{Iaconis} for details of a numerical implementation. 

\subsection{Scaling of the entropies and topological corrections}

For the ground state of two-dimensional gapped quantum systems, the entanglement entropy obeys an area law,  i.e. it grows as the length $\ell$ of the boundary of system $A$ instead of its volume 
\begin{align}
S_{A} & =c_\ell \ell-\gamma_{\rm quantum}+\ldots\, .
\label{Eq:AreaLaw}
\end{align}
The coefficient $c_\ell$ is non-universal, as it depends on the microscopic details of the model and can be changed continuously.
More interesting is the presence of the subleading constant term $\gamma_{\rm quantum}$, which is often called topological entropy and is robust against changing the microscopic details of the system. 
For a smooth cut it does neither depend on any length scale of the system nor the geometry of the bipartition and as such must be rooted in long-range entanglement. 
The seminal work of Refs.~\cite{LevinWenEntropy,KitaevPreskill} showed that the topological entropy is universal and can be directly calculated from the total quantum dimension $\mathcal{D}$ of the effective topological field theory. In particular, it is given by $\gamma_{\rm quantum}=\log\mathcal{D}$.

\begin{figure}[b]
\includegraphics[width=\columnwidth]{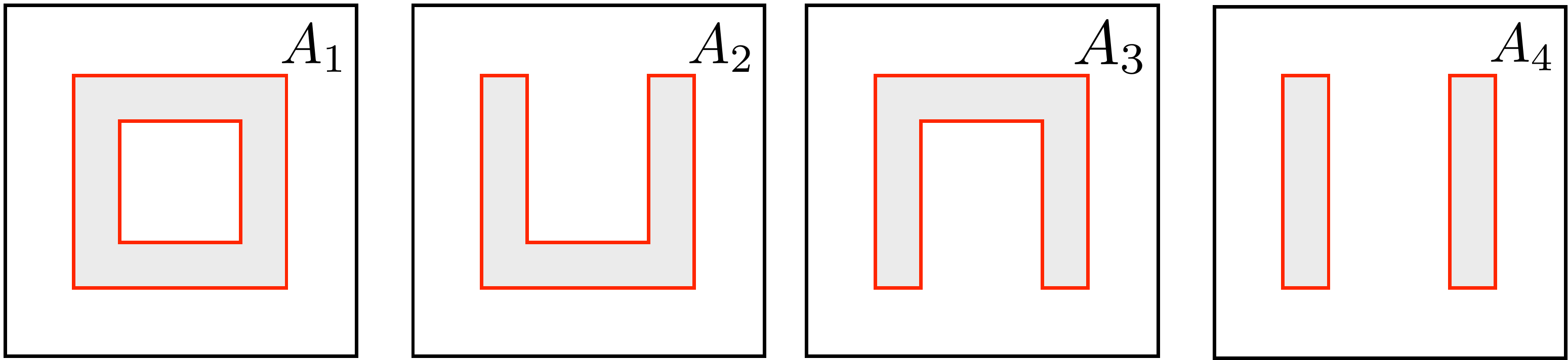}
\caption{(color online) Levin-Wen partitions to compute the topological entropy.  }
\label{fig:LevinWenCuts}
\end{figure}

As the topological entropy is  subleading, it is often cumbersome to determine it to sufficient accuracy by doing a scaling analysis, though recent numerical investigations have been quite successful in doing so \cite{HongChenNature,Kagome}. 
In addition, there may be additional $O(1)$ contributions to $S_A$ arising from sharp corners in the cut. 
Thus, it is beneficial to use a setup, which was introduced by Levin and Wen \cite{LevinWenEntropy}, to extract the topological entropy from a set of different bipartitions of the system as illustrated in Fig.~\ref{fig:LevinWenCuts}. 
By considering the four different bipartitions of Fig.~\ref{fig:LevinWenCuts} one can cancel out the leading boundary term in 
Eq.~\eqref{Eq:AreaLaw} as well as all $O(1)$ potential corner contributions to directly obtain the topological entropy from the linear combination
\begin{align}
\label{eq:classical topological entropy}
S_{\rm topo} & =-S_{A_1}+S_{A_2}+S_{A_3}-S_{A_4} \,,
\end{align}
which then leads to the final result 
\begin{equation}
  S_{\rm topo}=2\gamma_{\rm quantum} \,.
\end{equation}

In contrast to the quantum case, the leading term of the classical Shannon entropy $S_A$ scales with the volume $V_A$ of the subsystem $A$ in the bipartition, i.e. it follows a volume law of the general form
\begin{align}
S_{A} & =c_{V}V_{A}+c_{\ell}\ell-\gamma_{\rm classical}+\ldots \,,
\label{eq:general form of partition entropy}
\end{align}
where $\ell$ is again the length of the boundary of the partition and the  dots indicate subleading terms of order $O(1/\ell$) that vanish in the limit of $V_A, \ell\rightarrow \infty$. Note again that a direct consequence of this volume law is that the Shannon entropy is in general not symmetric with regard to the two partitions, i.e. $S_A \neq S_B$.
Similar to the quantum case, we can also in the classical system identify a constant contribution $\gamma_{\rm classical}$, which reduces the entropy. This constant turns out to be independent of the geometry and size of the system, which is why we will call it `classical topological entropy'. As we will discuss in the subsequent sections we can uniquely determine the universal values for this classical topological entropy for a broad variety of classical topologically ordered systems.

\begin{figure}[t]
\includegraphics[width=\columnwidth]{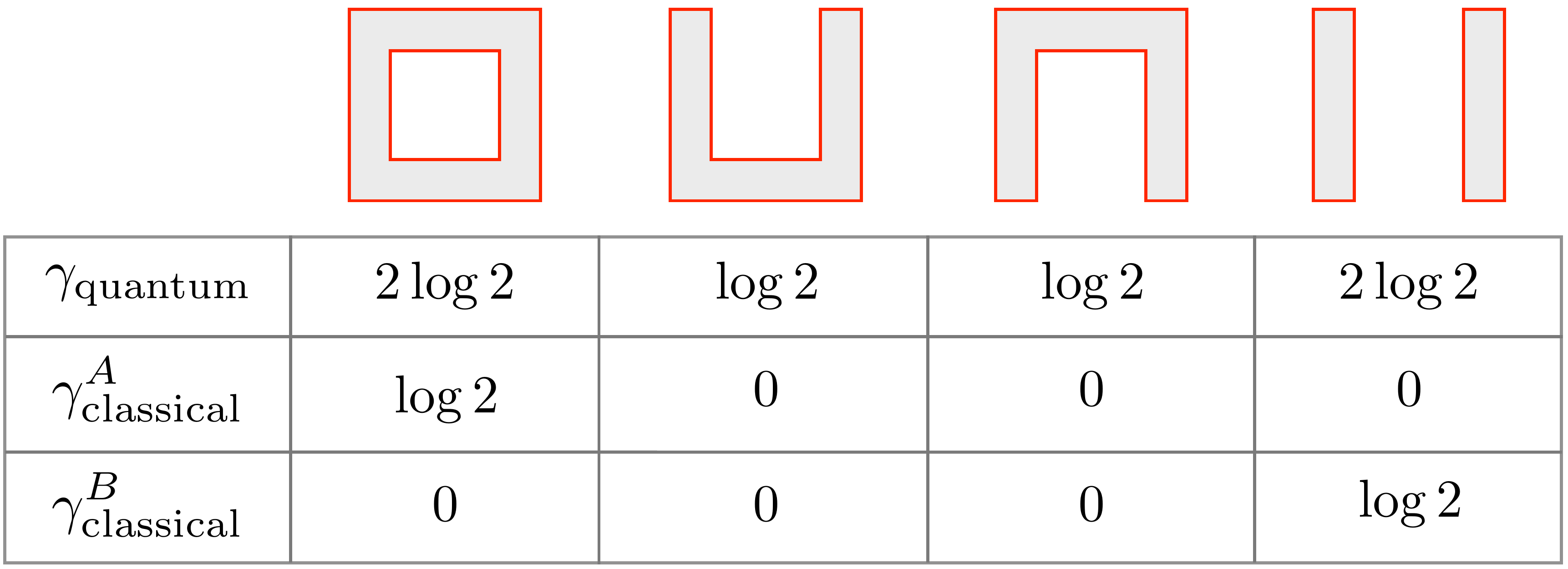}
\caption{(color online) Levin-Wen partitions to compute the topological entropy. Below the partitions is the constant term that is contributed by this bipartitions for the toric code ($\gamma_{\rm quantum}$) and the classical loopgas, when computing $S_A$ ($\gamma_{cl}^A$) respectively $S_B$ ($\gamma_{cl}^B$). }
\label{fig:LevinWenPartitions}
\end{figure}

At this point, we only want to point out that the form of the scaling behavior of the classical Shannon (and Renyi) entropies still allows to use the setup suggested by Levin and Wen to directly evaluate the $O(1)$ topological correction.
One subtle difference to the quantum mechanical calculation is that in the classical case the topological correction $\gamma_{\rm classical}$ is sensitive to the number of disconnected regions in partition $B$.
As an example, we show the constant contributions to the entropy for a quantum system and its related classical system for the four Levin-Wen partitions in Fig.~\ref{fig:LevinWenPartitions}.  The first line indicates the topological entanglement entropy for the
quantum mechanical system, the example being the so-called toric code model \cite{ToricCode}, which has total quantum dimension $\mathcal{D}=2$. 
Note that in the quantum mechanical case the topological correction for a given bipartition is related to the number of boundaries between parts $A$ and $B$; the bipartitions involving $A_1 $ and $A_4$ contribute twice the value than those bipartitions involving $A_2$ and $A_3$. 
The classical analog of the toric code is the so-called loopgas at $T=\infty$, whose classical topological entropy was already studied in Ref.~\cite{CastelnovoChamon1} and found to be $\gamma_{\rm classical}=\log 2$. As the Shannon entropy is not symmetric in $A$ and $B$, we show both the topological correction when computing $S_A$ (in the second line) as well as $S_B$ (in the third line). Note that the classical value is independent of the number of boundaries but instead measures the number of disconnected regions in the complementary part. For instance, the topological $O(1)$ contribution  in $S_A$ is non-vanishing only if part $B$ is disconnected.

Finally note that the similar setup introduced by Kitaev and Preskill \cite{KitaevPreskill} for extracting the topological entropy is not suitable for the classical systems considered here, because all bi-partitions have connected parts $A$ and $B$.


\section{Classical stringnet models}
\label{sec:classical_stringnet_models}

As an example family of model systems exhibiting non-trivial topological order, we will consider so-called stringnet states and in particular their classical analogs. Their quantum variants, so-called stringnet condensed states or simply stringnets, have been introduced by Levin and Wen \cite{LevinWenStringnet} in a rather general mathematical construction using so-called quantum doubles \cite{DrinfeldDoubles,QuantumDoubles}. For completeness, we will briefly recap this Levin-Wen construction in the following by first introducing so-called su(2)$_k$ anyon theories and then outlining the quantum double construction before shifting gear to discuss the construction of their classical analogs.

\subsection{su(2)$_k$ anyon theories}

The most elementary building block for both the quantum and classical versions of stringnets are so-called su(2)$_k$ anyon theories \cite{Wang}. These theories describe anyonic degrees of freedom with both Abelian and non-Abelian exchange statistics (in two spatial dimensions). For a given level $k$ such an anyonic theory contains $k+1$ individual degrees of freedom, which for our purposes here we label by integers
\[
  0, 1, 2, \ldots k \,.
\]
One can think of every degree of freedom as a representation of the quantum group su(2)$_k$ or in analogy to the spin representation of SU(2), which corresponds to the limit $k \to \infty$, as generalized spins with even/odd integer labels corresponding to integer/half-integer spins.

We can combine two representations of su(2)$_k$ into one joint representation -- similar to combining two spin quantum numbers into one joint spin quantum number for conventional SU(2) spins. This process, which for the anyon theories is often called fusion, has to obey very similar rules as those for combining two conventional SU(2) spins. In particular, they have to obey the so-called fusion rules which also incorporate the cut-off $k$ in a consistent way
\begin{align}
\label{eq:fusion rules}
i\times j & =\sum_{l=|i-j|}^{\min[i+j,2k-i-j]}l\, ,
\end{align}
where $l$ increases in steps of two. Eq.~\eqref{eq:fusion rules} can be written more compactly by introducing fusion coefficients $\mathcal{N}_{ij}^{l}$, which are defined via
\begin{align}
\label{eq:fusion_coefficients}
i\times j & \equiv \sum_{l=0}^{k}\mathcal{N}_{ij}^{l}\, l\,.
\end{align}
Note that for the su(2)$_k$ anyon theories at hand these fusion coefficients are always either 0 or 1.

The simplest example is probably the anyon theory su(2)$_1$ with anyonic degrees of freedom $0$ and $1$, 
for which the above fusion rules become
 \begin{align}
 \label{eq:fusion rules su(2)1}
 0\times 0&=0\nonumber\\
 0\times 1&=1\nonumber\\
 1\times 1&=0 \,.
 \end{align}
 A slightly less trivial example is the anyon theory su(2)$_2$ with anyonic degrees of freedom $0,1$ and $2$,
for which the fusion rules read
 \begin{align}
 \label{eq:fusion rules su(2)2}
 1\times 1&=0+2\nonumber\\
 1\times 2&=1\nonumber\\
 2\times 2&=0\, ,
 \end{align}
where the fusion with the identity $0$ has been omitted, as it is trivial.  
In addition we also mention the fusion rules of su(2)$_3$ with anyonic degrees of freedom $0,1,2$ and $3$,
for which the fusion rules read
\begin{align}
\label{eq:su(2)3}
1\times 1&=0+2\nonumber\\
1\times 2&=1+3\nonumber\\
1\times 3&=2\nonumber\\
2\times 2&=0+2\nonumber\\
2\times 3&=1\nonumber\\
3\times 3&=0 \,
\end{align}
and finally the Fibonacci theory, which is the even-integer subset of this su(2)$_3$ anyonic theory
\begin{align}
\label{eq:Fibonacci}
0\times 0&=0\nonumber\\
0\times 2&=2\nonumber\\
2\times 2&=0+2 \,.
\end{align} 

One striking distinction between the fusion rules for su(2)$_1$ in \eqref{eq:fusion rules su(2)1}  and for the remaining ones in \eqref{eq:fusion rules su(2)2}, \eqref{eq:su(2)3}, and \eqref{eq:Fibonacci}  is the occurrence of representations, which fused with itself, generate more than one fusion outcome, e.g. representation $1$ in su(2)$_2$. Such representations are called {\em non-Abelian} as opposed to {\em Abelian} representations such as the identity $0$ that always generate a unique fusion outcome \cite{FootnoteNonAbelianess}.  
In order to understand this concept better, let us consider a set comprising multiple such non-Abelian representations. Their combined fusion will no longer be described by a single state but necessarily needs to be described by a set of states. In more technical terms, this manifold of states for a set of non-Abelian degrees of freedom asymptotically grows exponentially with the total number of degrees of freedom. The base of this exponential growth is called the {\em quantum dimension} $d_j$ of the representation $j$. Non-Abelian representations have quantum dimensions that are strictly larger than one, i.e. $d_j > 1$. In contrast, any representation $i$ that is Abelian (exhibiting only single fusion outcomes) has quantum dimension $d_i = 1$.
For a given anyon theory the {\em total quantum dimension} $\mathcal{D}$ is then defined as a sum over all the quantum dimensions of the individual representations of the theory
\begin{align}
\label{eq:tot qu dim}
\mathcal{D}&=\sqrt{ \sum_{j=0}^k d_j^2 }\, .
\end{align}
For all values of $k$, the su(2)$_k$ anyon theories always contain two Abelian representations, labeled by $0$ and $k$, 
with quantum dimensions $d_0=d_k=1$. The remaining representations are all non-Abelian and thus have quantum dimension $d_j>1$.

\subsection{Quantum double models and stringnet condensed states}

\begin{figure}[b]
\includegraphics[width=\columnwidth]{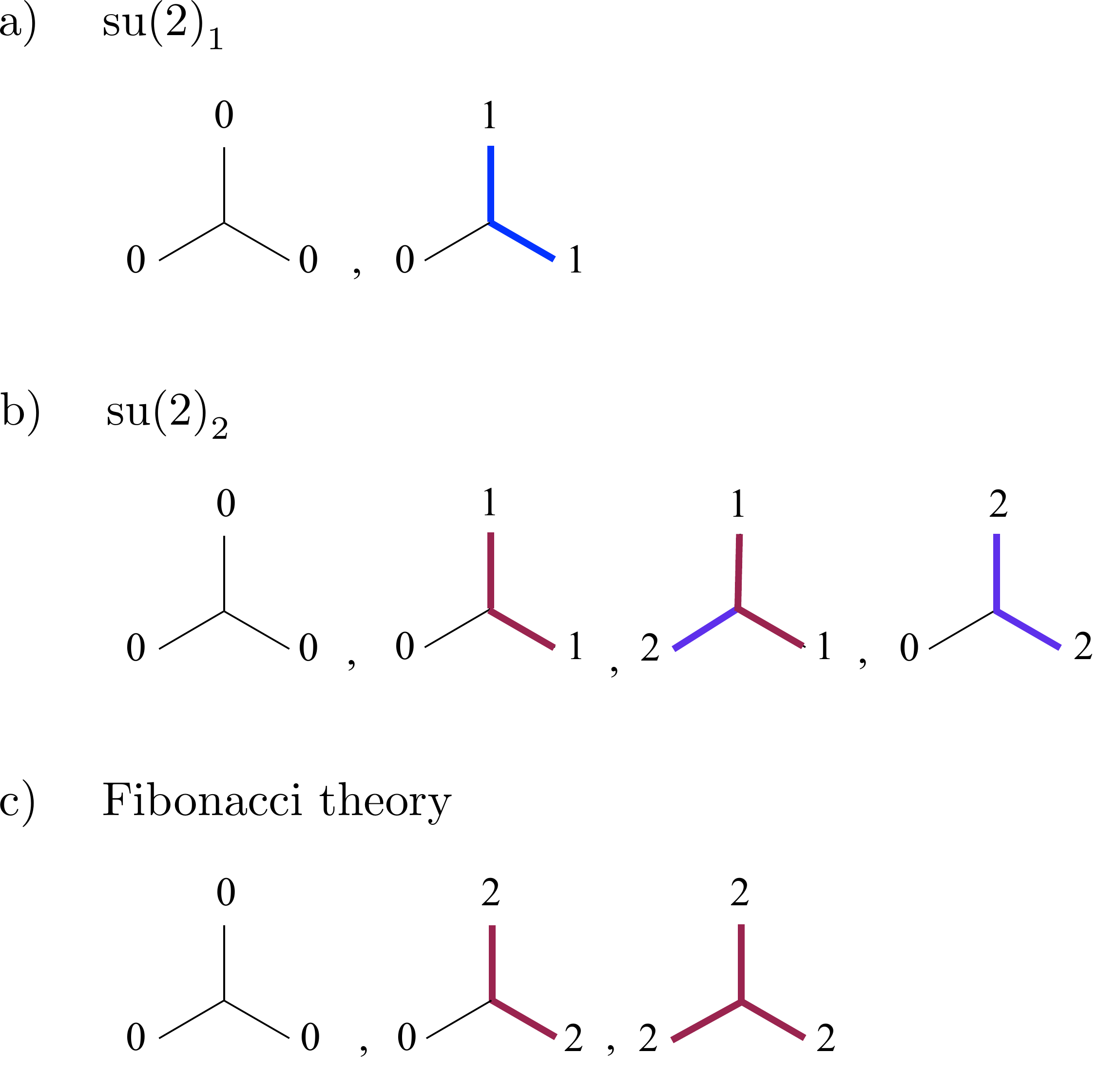}
\caption{(color online) Allowed vertices (up to rotations) for a) the su(2)$_1$ anyon theory, b) the su(2)$_2$ anyon theory, and c) the Fibonacci theory.}
\label{fig:allowed_vertices}
\end{figure}

With the su(2)$_k$ anyon theories as elementary building blocks at our hand, we can now proceed to briefly recap the quantum double construction of Levin and Wen \cite{LevinWenStringnet} and introduce the quantum mechanical version of stringnet condensed states or simply stringnets.

The quantum double construction of Levin and Wen creates a lattice model from an anyonic theory, such as one of the su(2)$_k$ anyon theories introduced in the previous section. The elementary constituents of the lattice model are edges that carry an anyonic degree of freedom, which is captured by the respective anyon theory, i.e. it corresponds to one of the labels $0,1,2, \ldots,k$. The bonds form a lattice via {\em trivalent} vertices. An example of a particularly regular lattice construction would be the honeycomb lattice, which we will use in all visualizations in the following.
At each vertex the fusion rules of the anyonic theory need to be fulfilled, thus constraining the possible labelings of edges around a given vertex. Examples of allowed vertices for the su(2)$_1$, su(2)$_2$ and Fibonacci theory are given (up to rotations) in the various panels of  Fig.~\ref{fig:allowed_vertices}, respectively.

\begin{figure}[t]
  \begin{centering}
  \includegraphics[width=.65\columnwidth]{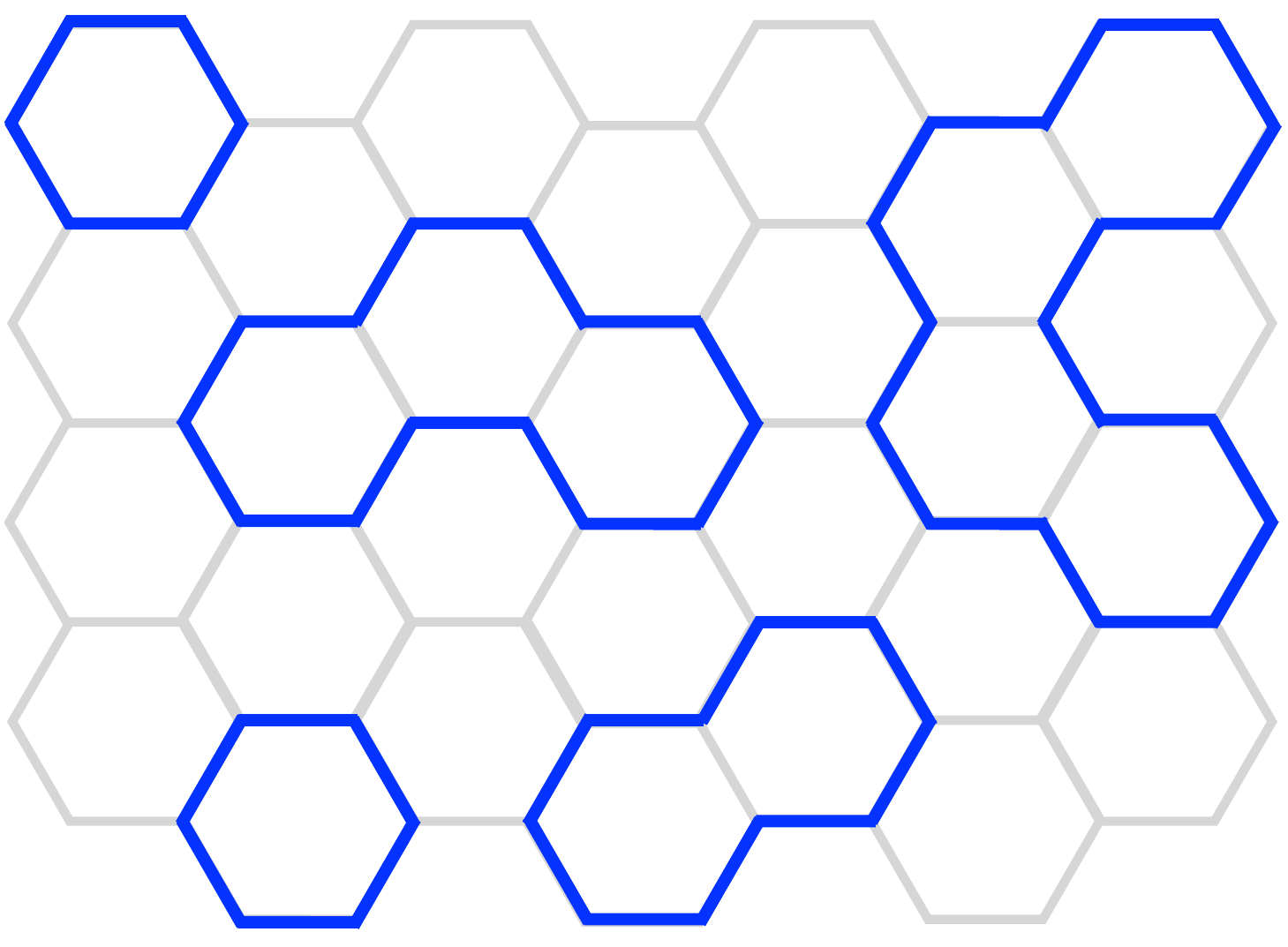}
  \end{centering}
  \caption{(color online) Example loopgas configuration for the quantum double model of su(2)$_1$.}
  \label{Fig:LoopGas}
\end{figure}

\begin{figure}
  \begin{centering}
  \includegraphics[width=.65\columnwidth]{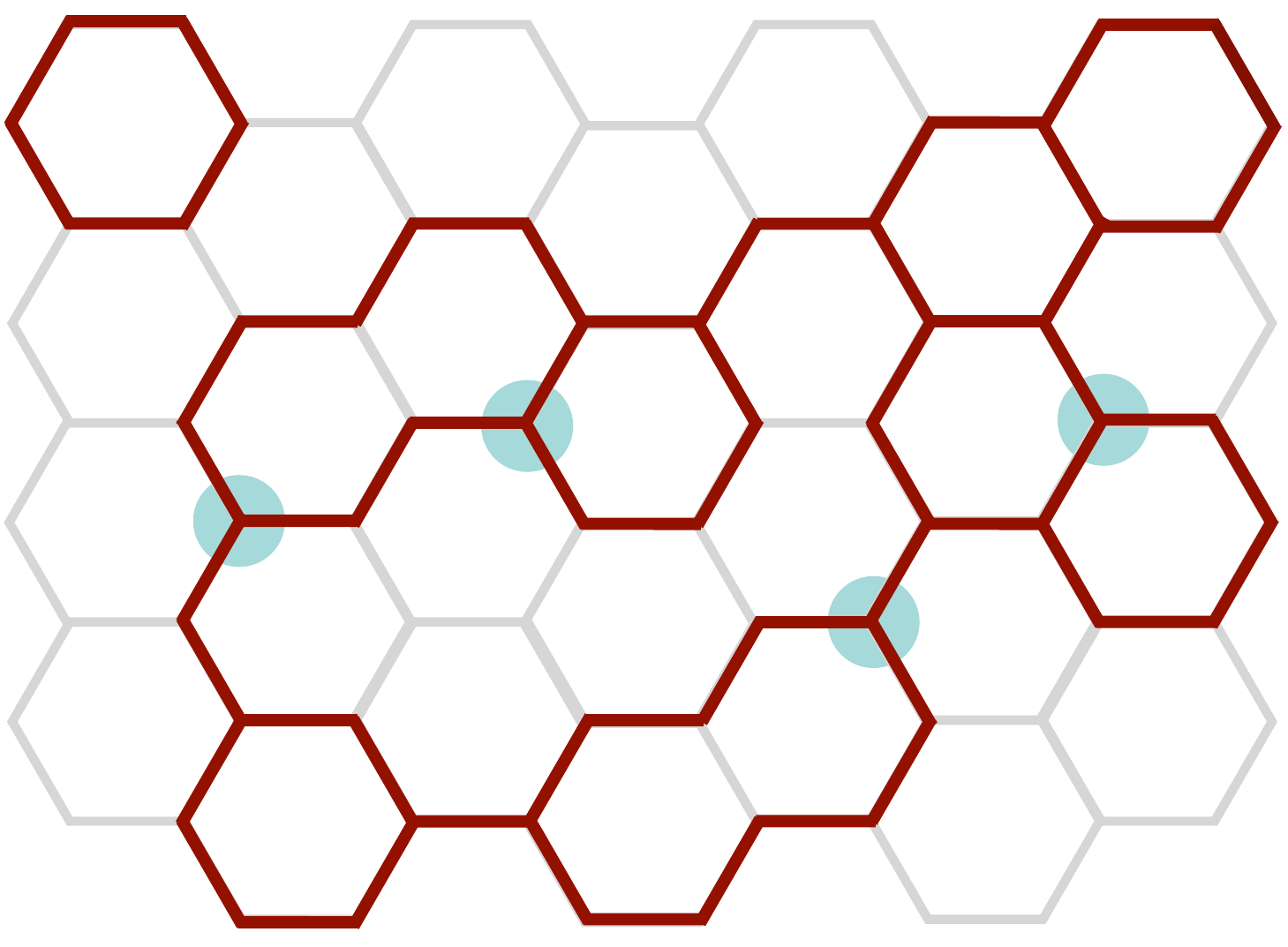}
  \end{centering}
  \caption{(color online) Example stringnet configuration for the quantum double model of the Fibonacci theory. 
  		The circles indicate some of the branching points not present in the loopgas configuration of Fig.~\ref{Fig:LoopGas}
		for the Abelian su(2)$_1$ anyon theory.}
  \label{Fig:StringNet}
\end{figure}

Forming an entire graph out of a set of these allowed vertices and corresponding edges, one can then create a multitude of lattice configurations.
By visual inspection of these configurations it becomes evident how the vertex constraints translate into constraints of the overall lattice configurations. For the case of the su(2)$_1$ anyonic theory, which contains only Abelian anyons, the resulting configurations will consist of closed loops only, as illustrated in Fig.~\ref{Fig:LoopGas}. Introducing a non-Abelian anyon into the anyon theory, as it is the case for, e.g. the Fibonacci theory, the resulting lattice configurations also include branches resulting in netlike configurations as illustrated in Fig.~\ref{Fig:StringNet}.

\begin{figure}
  \begin{centering}
  \includegraphics[width=.65\columnwidth]{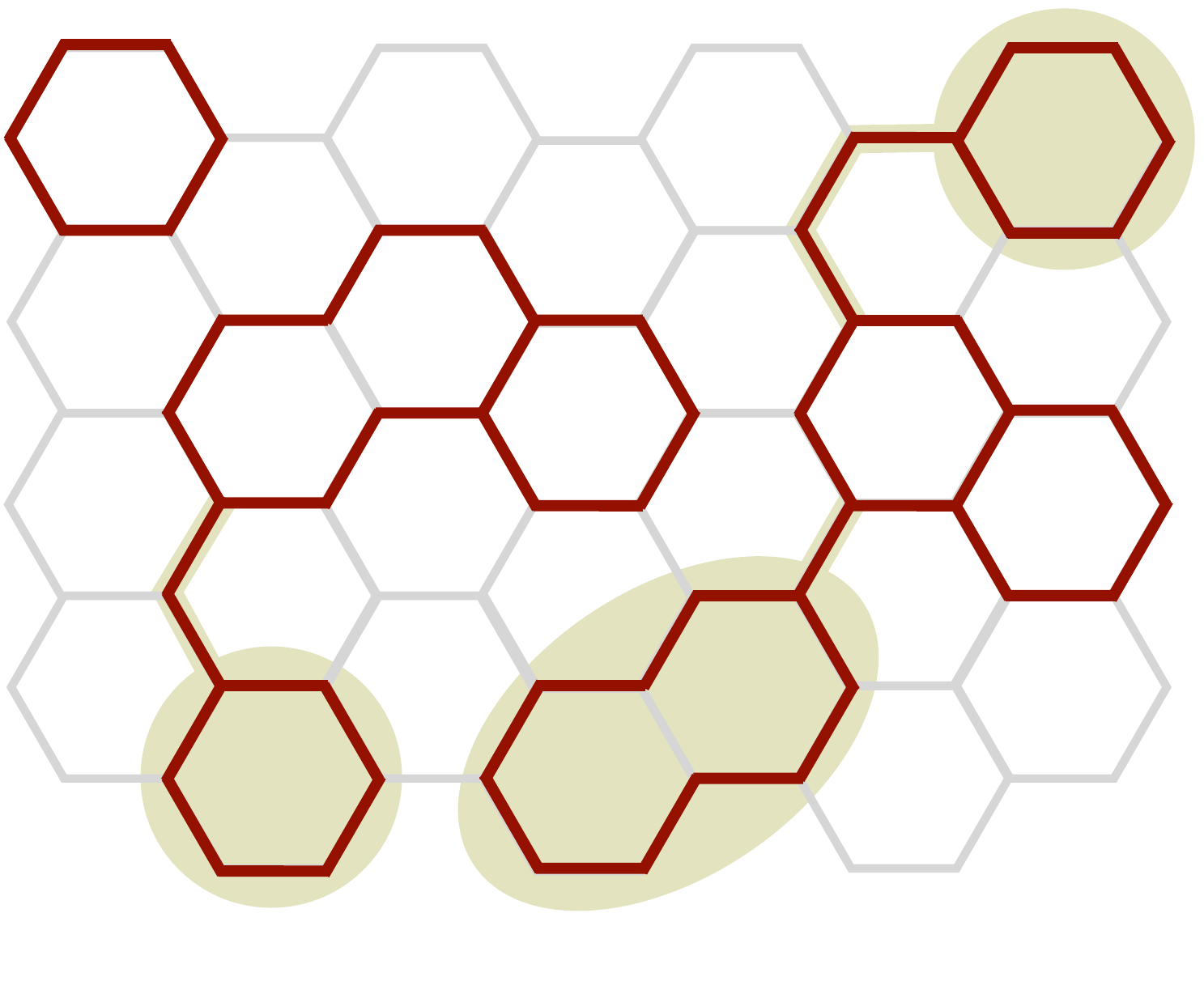}
  \end{centering}
  \caption{(color online) Example stringnet configuration which includes so-called tadpoles indicated by the shaded ovals.}
  \label{Fig:StringNetTadpole}
\end{figure}

The set of allowed lattice configurations forms a basis for the part of the Hilbert space, where the groundstate resides in. Before moving on to the construction of the classical analog of these quantum states, we want to mention in passing a few additional constraints present in the quantum model (but not the classical model to be introduced in the next section). Most notable for the remainder of this manuscript is that  a subset of allowed lattice configurations has vanishing weight for the groundstate, in particular all those  configurations where the netlike configuration includes so-called tadpoles illustrated in Fig.~\ref{Fig:StringNetTadpole}. The actual weight of an allowed lattice configuration will be finely tuned by a number of parameters depending on the specifics of the underlying anyon theory (such as, e.g., a $d$-isotopy parameter for the inclusion of closed loops). Further, the Levin-Wen construction goes beyond the construction of quantum ground states, but also allows a description of excited states, e.g. states where the vertex constraint is not fulfilled at individual vertices or which include one of the aforementioned tadpoles.

Finally we note that in more mathematical terms, the quantum double construction is routed in the so-called Drinfeld double \cite{DrinfeldDoubles},
which takes a topological quantum field theory (TQFT) $\mathcal{C}$ and its time-reversal conjugate $\mathcal{C}^*$ to construct a doubled TQFT that is time-reversal invariant \cite{Mueger}. For more details on this construction the reader is referred to the original math literature \cite{DrinfeldDoubles,Mueger} or the more physical introduction in Ref.~\cite{QuantumDoubles}. Note that the total quantum dimension of the quantum double model is the square of the total quantum dimension of the anyonic theory, i.e. $\mathcal{D}_{\mbox{\tiny QDM}}=\mathcal {D}^2 =\sum_{i=0}^k d_i^2$, resulting in a topological entropy $\gamma_{\mbox{\tiny QDM}}=\log \mathcal D ^2$.

\subsection{Classical stringnets}
\label{sec:classical_stringnets}
 
We now turn to the classical analogs of the quantum mechanical stringnet states introduced as groundstates of the quantum 
double construction of Levin and Wen in the previous section. To do so, we first note that the quantum ground states of Levin and Wen can generally be written as a superposition of stringnet configurations $s$, which fulfill the vertex constraints at every single vertex. The wavefunction can then be expressed as
\begin{align}
   | \psi \rangle = \sum_s a_s | s \rangle \,,
\end{align}
where the coefficients $a_s$ are real, but not necessarily all positive or non-zero \cite{FootnoteToricCode}. 

We can now construct a {\em classical} stringnet by defining a partition function that corresponds to the equal-weight superposition of allowed stringnet configurations $s$ of the quantum double construction, i.e.
\begin{align}
  Z[V] = \sum_s  1 \,,
  \label{eq:partition_function}
\end{align}
where $V$ indicates the number of vertices in the lattice.
Note  that whether a given stringnet configuration $s$ is allowed or not, again depends solely on the {\em local} constraints implemented around every vertex. In fact, one can explicitly count the number of allowed configurations contributing to the partition function \eqref{eq:partition_function}. Going through a sequence of combinatorial steps (detailed in Appendix \ref{sec:stringnet}), 
one finds that 
\begin{align}
\label{eq:Z_approx}
Z[V]=\sum_{j=0}^k \left(\frac{\mathcal{D}}{d_j}\right)^V \approx 2\mathcal{D}^V \,,
\end{align}
where the $d_j$ indicate the quantum dimension of the anyons in the underlying anyon theory su(2)$_k$ and $\mathcal{D}$ is the total quantum dimension of the theory. In the large-volume limit, the explicit sum can be approximated as a power of the total quantum dimension, which connects the original meaning of the quantum dimension in the anyonic theory with its analogous role in the purely classical model.
 
Before we turn to a general discussion of the classical stringnets defined in this section, we will briefly layout the more technical details of the methods and in particular the role of a so-called crossing symmetry for the analytical calculations of the Renyi entropies for these classical stringnets.


\section{Crossing symmetry}
\label{sec:crossing_symmetry}

In the following we derive several formulas, which are important to the computation of the Shannon or Renyi entropies of the classical stringnet models.  Our analytical approach is heavily based on the concept of a so-called crossing symmetry, which was recently introduced by Fendley and Simon as a method to compute the exact partition function of a special class of classical lattice model \cite{SimonFendley}. For later convenience, we will keep the discussion in this section very general, such that it is applicable to {\em any} crossing symmetric model. As the following discussion of crossing symmetric models is very brief, the interested reader is referred to  Ref.~\cite{SimonFendley} for a more complete treatment. 

\subsection{Introduction to crossing symmetry}
In the following we consider  models that are defined on a trivalent graph with $N$ possible states -- labeled by $0,1,2,\ldots N-1$--- living on the edges of the graph. The edges are in general oriented, which is indicated by an arrow in the visualizations. We assign each degree of freedom a conjugate via a permutation $P$ of $[0,\ldots, N-1]$ of maximally order two, i.e. $P^2=1$. The conjugate of $i$ is denoted by $\bar i$ and conjugation is implemented by reversing the arrow on the edge. The physical significance of the conjugation will become clear below.

\begin{figure}
\includegraphics[width=0.8\columnwidth]{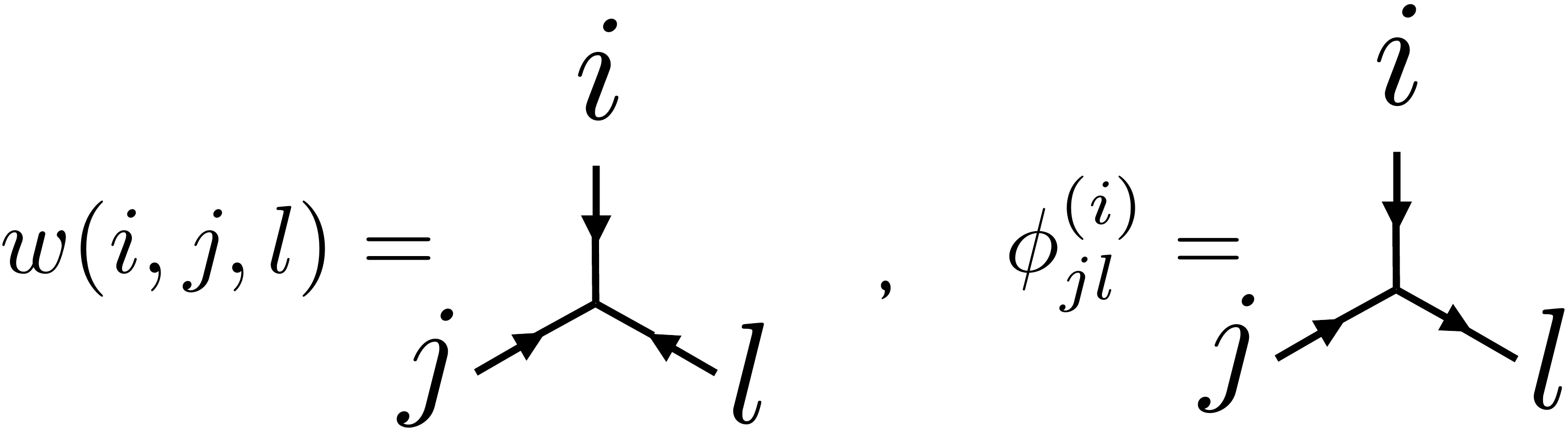}
\caption{Definition of the local Boltzmann weight of a vertex configuration.} 
\label{fig:weight_directed}
\end{figure}

We define local Boltzmann weights for the vertices that depend on the labels (and orientations) of the three edges around the vertex. When defining the weights $w(i,j,l)$ we use the reference orientation that all arrows are pointing inwards, as is visualized in Fig.~\ref{fig:weight_directed}. The vertex weights can alternatively be written as a `weight matrix' $\phi^{(i)}$ defined as
\begin{align}
\label{eq:weight matrix}
\phi^{(i)}_{i,j}=w(i,j,\bar l)\,, 
\end{align}
which will become convenient later on. Note that the second matrix index is conjugated compared to the definition of the Boltzmann weights, see also Fig.~\ref{fig:weight_directed}.  We consider only models that are isotropic, which implies that the value of the Boltzmann weight is the same for cyclic permutations of the indices,  i.e. $w(i,j,l)=w(j,l,i)=w(l,i,j)$. 
The partition function $Z[V]$ of a graph with $V$ vertices  is then defined as the sum over all edge labelings of the product of all weights 
\begin{align}
Z[V]&= \sum_{\mbox{\small edge labels}} \prod_{\mbox{\small vertices }\nu} w(i_\nu,j_\nu,l_\nu)\,.
\end{align}

\begin{figure}[tbh]
\centering{}
\includegraphics[width=0.7\columnwidth]{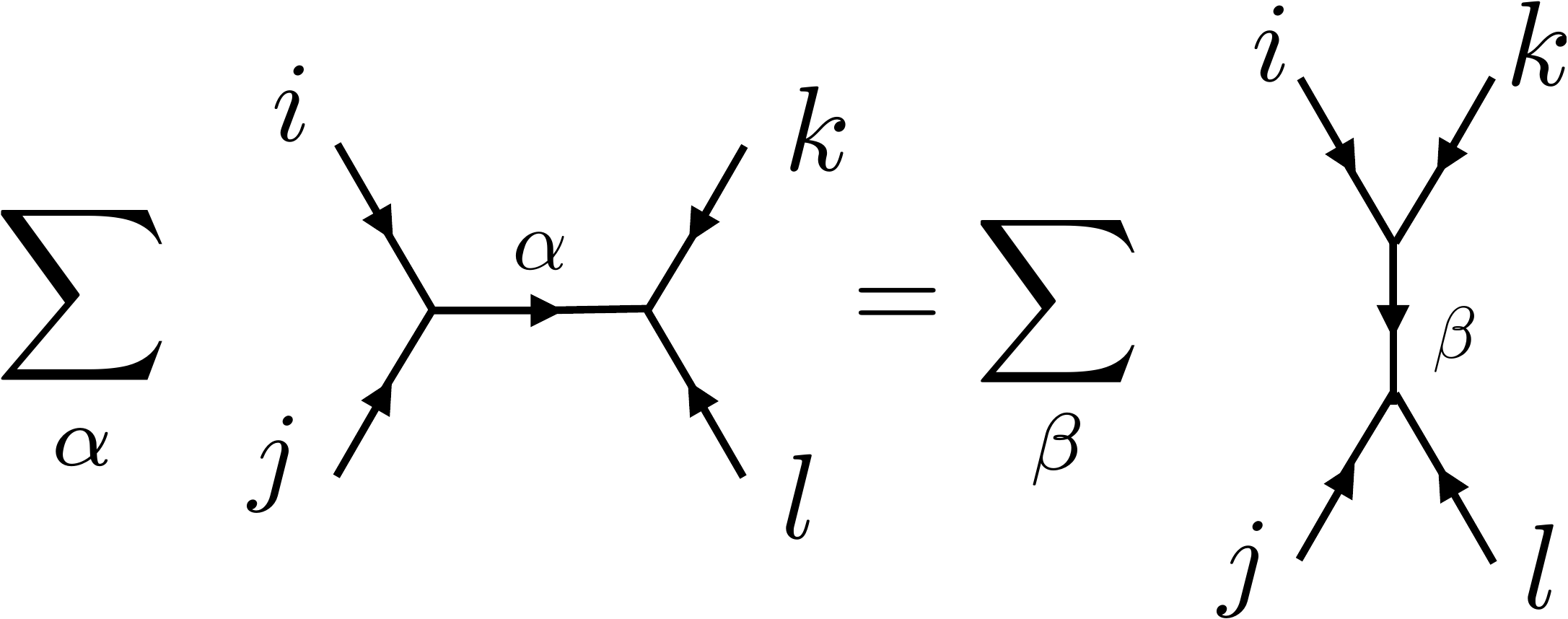}\caption{Graphical representation of crossing symmetry.  
}\label{fig:crossing symmetry}
\end{figure}

\begin{figure}[tbh]
\includegraphics[width=0.3\textwidth]{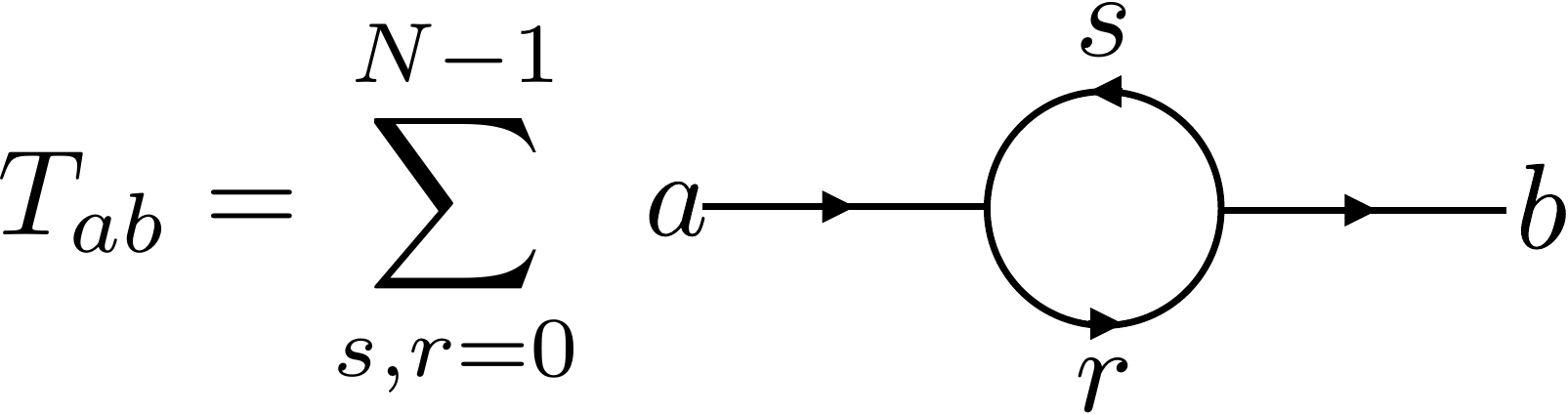}
\caption{Graphical representation of the matrix elements $T_{ab}$. }
\label{fig:T-matrix}
\end{figure}

The model is called crossing symmetric if the Boltzmann weights defined on the vertices fulfill the crossing symmetry relation 
\begin{align}
\sum_{\alpha=0}^{N-1}w(i,j, \bar\alpha)w(\alpha,l,k) & =\sum_{\beta=0}^{N-1}w(k,i, \bar\beta)w(\beta,j,l)\,, 
\label{eq:crossing symmetry}
\end{align}
which is visualized in Fig.~\ref{fig:crossing symmetry}. 
Crossing symmetry implies that the weight matrices \eqref{eq:weight matrix} commute. 
\begin{figure}[tb]
\includegraphics[width=\columnwidth]{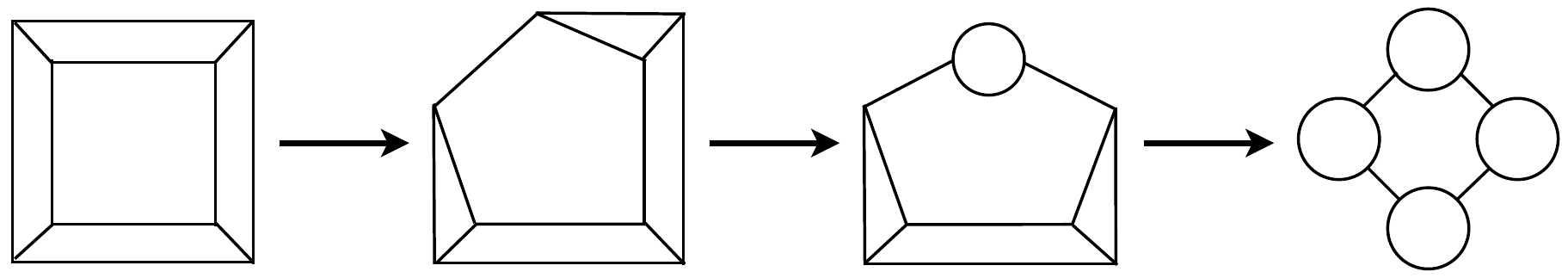}
\caption{Using crossing symmetry equates the partition function of different graphs. Any planar, closed graph can be reduced to a chain of bubbles (rightmost graph). }
\label{fig:crossing symmetry Z} 
\end{figure}
In Ref.~\cite{SimonFendley} Fendley and Simon showed that crossing symmetry allows for calculating the partition function exactly.  Using the crossing symmetry relation \eqref{eq:crossing symmetry} successively, one can equate the partition function of different graphs. This in turn allows a transformation to a graph, where the partition function can be calculated straightforwardly. An example of such a transformation is illustrated in Fig.~\ref{fig:crossing symmetry Z}. 
The main insight is that the graph can always be deformed into a chain of `bubbles', where a bubble can be interpreted as a matrix that in the following is denoted by  $T$ and defined through the weight matrices \eqref{eq:weight matrix}
\begin{align}
T & =\sum_{s=0}^{N-1}\phi^{(s)}\phi^{(\bar s)}\,.
\label{eq:T matrix}
\end{align}
By definition, the $T$-matrix commutes with all the weight matrices. 
The graphical interpretation of the bubble as a matrix is visualized in Fig.~\ref{fig:T-matrix}. 
As a result of the transformation of the graph, the evaluation of the partition function is reduced to diagonalizing an $N\times N$ matrix
\begin{align}
Z[V]&= \mbox{Tr}\left(T^{V/2}\right)\,. 
\label{eq:partition function}
\end{align} 

\subsection{Exampels of crossing symmetric models}

The crossing symmetry relation \eqref{eq:crossing symmetry} seems at first glance rather restrictive. However, there are a number of interesting classical models that obey the relation. In particular, it was shown in Ref.~\cite{SimonFendley} that the classical stringnet models discussed in Section \ref{sec:classical_stringnets} are crossing symmetric. To be more precise, the degrees of freedom living on the edges are the representations of the su(2)$_k$ anyon theory.  Admissible stringnet configurations have the property that the fusion rules are fulfilled at each vertex. This constraint corresponds to a local Boltzmann weight of  \begin{align}
 \label{eq:def_weight_su(2)k}
w(i,j,  l) & =\mathcal{N}_{ij}^{\bar l}\,, 
 \end{align}
where the ordering of the labels is not important, i.e.
 \[
 w(i,j, l) =w(j,i,l)\,. 
 \]
The conjugate $\bar i$ of the degree of freedom $i$ is defined as the unique element, such that the fusion of $i$ and its conjugate contains the identity element $0$, i.e. $i\times\bar i =0+\ldots$ . From the fusion rules of the su(2)$_k$ anyon theory,  it can readily be seen that the identity element only occurs when $\bar i=i$. Hence, all degrees of freedom are their own conjugate in this case and we can consider the edges as un-oriented. However,  this is no longer true for more complicated anyon theories, such as e.g.  su(N)$_k$. 

Another class of crossing symmetric models are the ones based on finite groups. The degrees of freedom living on the edges are now taken to be elements of a finite group $G$ with the identity element denoted by $e$. In this manuscript, we are most interested in models, where the group operation $\circ$ is fulfilled at each vertex, i.e. 
\begin{align}\label{eq:def_weight_group}
w(i,j,l)=\left\{
\begin{array}{cc}
1 & \mbox{if } i\circ j\circ l=e\\
0 &\mbox{otherwise}\,.
\end{array} \right.
\end{align}
The conjugate element $\bar i$ of $i$ is defined as the unique inverse, such that $i \circ \bar i=\bar i \circ i=e$.  When $G$ is non-Abelian the ordering of the indices in $w(i,j,l)$ is important, i.e. in general $w(i,j,l)\neq w(i,l, j)$. As a result, we need to introduce an orientation for each vertex (in addition to the orientation of the edges). Here, we use the convention that the ordering of the indices corresponds to an anti-clockwise orientation of the vertex.   

We should comment on that Eq.~\eqref{eq:def_weight_group} is, in fact, much more restrictive than is needed for crossing symmetry, as was shown in Ref.~\cite{SimonFendley}. However,  the models defined by \eqref{eq:def_weight_group} are the most interesting from the perspective of calculating the Renyi entropies, as it turns out that they have the maximal possible value of the topological entropy, namely $\log |G|$, where $|G|$ is the number of elements in $G$.

\subsection{Renyi entropy of crossing symmetric models}
\begin{figure}[tb]
\includegraphics[width = 0.4\columnwidth]{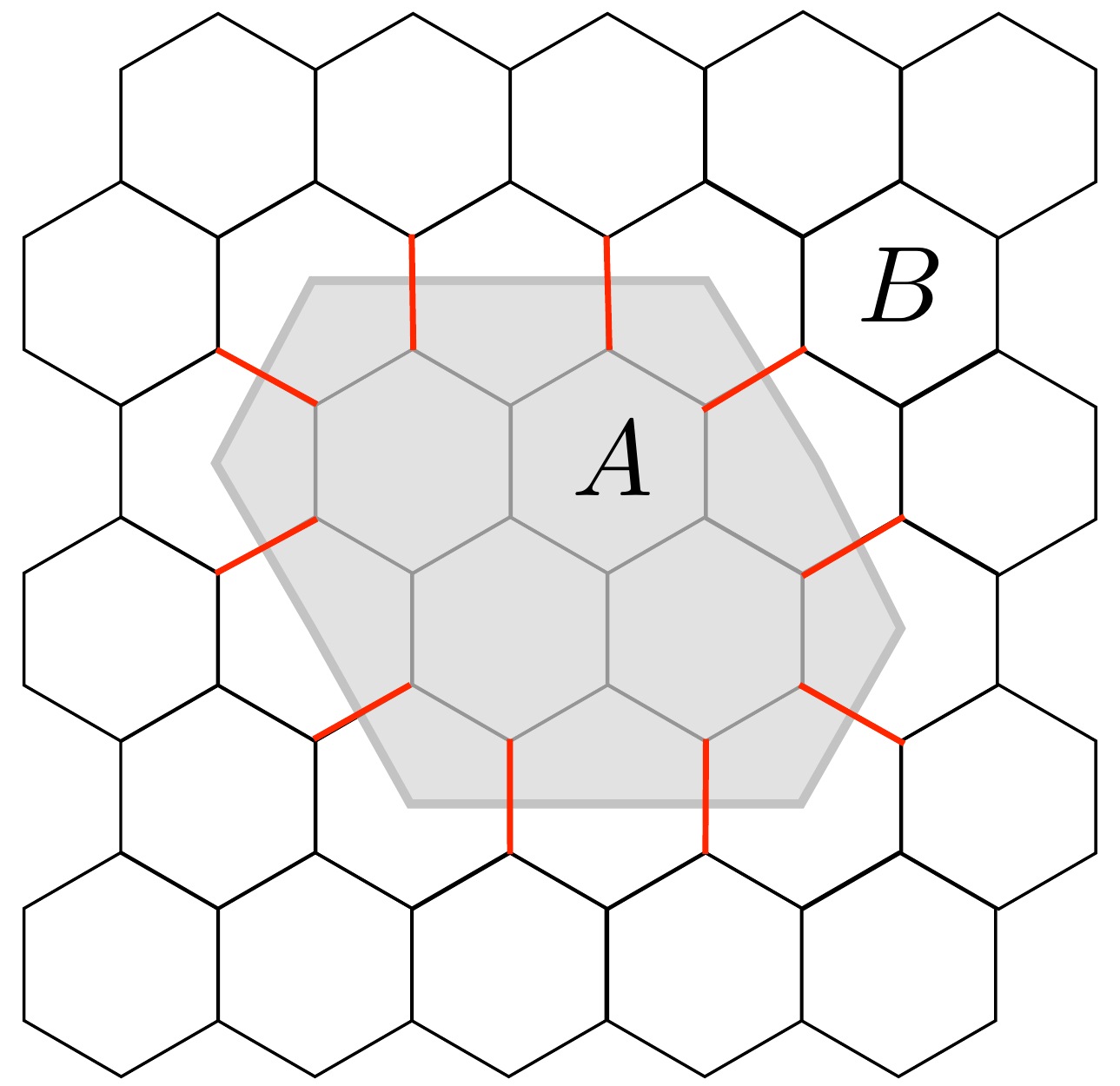}
\caption{(color online)  Bipartition of the lattice into two regions $A$ (in grey) and $B$. The system is cut on the edges, with the $\ell $ boundary edges marked in red. }
\label{fig:boundary}
\end{figure}

Having already derived an explicit expression for the partition function in Eq.~\eqref{eq:partition function}, let us now continue with discussing the relevant steps to compute the Renyi entropies. When dividing the system into two partitions $A$ and $B$, we divide the set of vertices spatially into vertices in $A$  and $B$ respectively, thus cutting the graph spatially on the edges. The total number of these `boundary edges' is denoted by $\ell$ in the following.   As the two subsystems $A$ and $B$ are only coupled via the boundary links,  it is useful to keep the boundary configuration, denoted by ${\boldsymbol \alpha} =(\alpha_1 ,\ldots, \alpha_\ell)$, explicit. A visualization of a bipartition of the system with boundary configuration ${\boldsymbol \alpha}$ can be found in Fig.~\ref{fig:boundary}.

In order to compute the Renyi entropies, we need an expression for the probability $p_{\{I ,\boldsymbol \alpha\}}$ of a given configuration  $\{I,\boldsymbol \alpha\}$, where the set $I$ incorporates all edge labels in the volume (excluding the boundary) of one of the partitions, say $A$. In terms of the local Boltzmann weights $w(i,j,l)$, the probability  is given by 
\begin{align}
p_{\{I ,\boldsymbol \alpha\}}&=\frac 1 Z \prod_{\nu\in A} w(i_\nu, j_\nu, l_\nu)\sum_{B}\prod_{\nu\in B} w(i_\nu,j_\nu,l_\nu)\,, 
\end{align}
where $\sum_B$ indicates a sum over all edge labels of links in $B$ (again excluding the boundary). Inserting the probability into the definition of the Renyi entropy with index $n$ yields
\begin{multline}
\label{eq:Renyi entropy general}
S_n^A =\frac 1 {1-n} \log\left[\sum_{\boldsymbol \alpha}\left(\sum_{A}\prod_{\nu\in A} w(i_\nu,j_\nu,l_\nu)^n \right)\right. \\
\left.\times \left(\frac{\sum_{B}\prod_{\nu\in B} w(i_\nu,j_\nu,l_\nu)}{Z}\right)^{n}\right] \,.
\end{multline}
In order to simplify this expression, we define a `boundary weight' $W_n(\boldsymbol \alpha, V_B)$ (and similarly for $A$) by
\begin{align}
\label{eq:Weightn}
W_n(\boldsymbol \alpha ,V_B)&=\sum_{B}\prod_{\nu\in B} w(i_\nu,j_\nu,l_\nu)^n\,, 
\end{align}
where  $V_B$ denotes the volume of $B$ and is defined such that $V_B+\ell$ is the total number of vertices in $B$. This definition ensures that the contributions of the volume and the boundary to the Renyi entropy are nicely separated in the final result.  

 $W_1(\boldsymbol \alpha ,V_B)$  is nothing but the total weight of classical configurations in $B$, given a particular boundary configuration ${\boldsymbol \alpha}$. 
\begin{figure}[bt]
\includegraphics[width = \columnwidth]{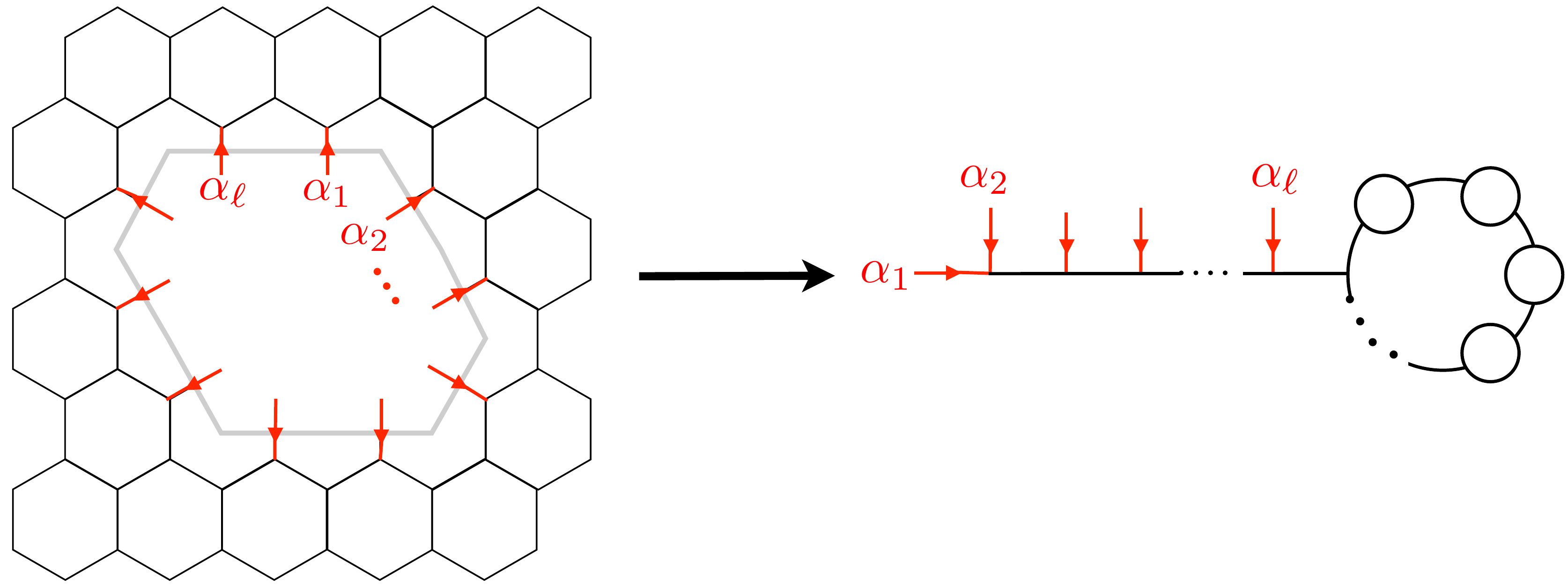}
\caption{(color online)  Graphical representation of how to use crossing symmetry to compute $W_1(\boldsymbol \alpha, V)$, Eq.~\eqref{eq:Weight}. The boundary is indicated in grey and the boundary edges as well as their orientation are marked in red. 
\label{fig:crossing symmetry boundary}}
\end{figure}
It can be computed exactly by using crossing symmetry to transform the graph of subsystem $B$ into one, where the internal summation over all edge labelings in $B$ is expressed as a matrix power of the $T$ matrix \eqref{eq:T matrix}:
\begin{align}
\label{eq:Weight}
W_1(\boldsymbol \alpha ,V_B)&=\sum_{\beta=0}^{M-1}\left[\phi^{(\alpha_2)} \ldots \phi^{(\alpha_\ell)} \right]_{\alpha_1,\beta}\mbox{Tr}\left(\phi^{(\beta)}T^{V_B/2}\right)\,. 
\end{align}
An example of such a transformation is illustrated in Fig.~\ref{fig:crossing symmetry boundary}. 
As the weight matrices $\phi^{(\alpha)}$ commute with each other as well as with the $T$-matrix, we can simultaneously diagonalize them and Eq.~\eqref{eq:Weight} can be readily expressed in terms of the eigenvalues of the matrices.  
Note that we must define a reference configuration on the boundary edges, in the case where  edges have an orientation. 
In the following, we will use the convention that the arrows point towards the vertices in the respective volume, i.e. $B$ in case at hand.  Consequently,  we need to take the conjugate boundary configuration $\bar{ \boldsymbol \alpha}=(\bar \alpha_1,\ldots,\bar \alpha_\ell)$ for the boundary weight of the other partition, which is  $A$ in this case. 

If the replicated system is crossing symmetric, i.e. if the weights $w(i,j,l)^n$ obey the crossing symmetry relation \eqref{eq:crossing symmetry}, we can also compute $W_n(\bar{ \boldsymbol \alpha} ,V_A)$ exactly, using the weight matrices and the $T$-matrix that are obtained from the weights $w(i,j,l)^n$ instead. As a result,  we can find a compact expression for the Renyi entropy of subsystem $A$ with index $n$ 
\begin{align}
\label{eq:Renyi entropy crossing symmetry}
S_n^A =\frac 1 {1-n} \log\left[\sum_{\boldsymbol \alpha}W_n(\bar{ \boldsymbol \alpha} ,V_A) \left(\frac {W_1(\boldsymbol\alpha,V_B)}{Z}\right)^n\right]\,,
\end{align}
for all classical models, where the model as well as the replicated model -- i.e. the model, where the local Boltzmann weights are given by $w(i,j,l)^n$ -- are crossing symmetric.  

The requirement that both $w(i,j,l)$ as well as $w(i,j,l)^n$ obey the crossing symmetry relation seems at first glance rather restrictive. However, we note that for the classical stringnet models based on the su(2)$_k$ anyon theories the weights $w(i,j,l)$ are all either $0$ or $1$. Hence, the replicated model is trivially crossing symmetric. The same applies to a few other stringnet models, e.g. those based on the su(N)$_k$ theories with $k=1,2$. It is also valid for the classical models based on finite groups that were discussed in the previous section.

%

\section{Renyi  entropy of classical stringnet models}
\label{sec:Renyi entropy}
In this section, we give a brief outline on how to derive the Shannon and Renyi entropies of classical stringnet models and discuss their properties. 
We focus on the model based on the su(2)$_k$ anyon theory, for which we provide the basic steps. A detailed derivation can be found in Appendix~\ref{sec:stringnet}. 

Let us first discuss  a bipartition, where part $A$ and $B$ are both connected, for instance $A_2$ and $A_3$ in Fig.~\ref{fig:LevinWenPartitions}. 
We note that for the classical stringnet based on the su(2)$_k$ anyon theory, all representations are self-conjugate and the local Boltzmann weight of a vertex is either $0$ or $1$. Consequently, the expression for the Renyi entropy \eqref{eq:Renyi entropy crossing symmetry} simplifies to the  following
\begin{align}
\label{eq:Renyi su(2)_k}
S_n^A  =\frac 1 {1-n} \log\left[\sum_{\boldsymbol \alpha} W_1(\boldsymbol \alpha,V_A) \left(\frac {W_1(\boldsymbol \alpha,V_B )}{Z}\right)^n\right]\, .
\end{align}
The partition function $Z[V]$, see Eq.~\eqref{eq:Z_approx}, can be approximated by 
\begin{align}
\label{eq:Zsu(2)}
Z[V]\approx 2\mathcal{D}^V
\end{align}
in the limit of large volume $V$. In the following, we assume that both the volume of partition $A$ and $B$ are large, such that the approximation above is valid. 
The boundary weights $W_1(\boldsymbol \alpha,V)$ can then be expressed in terms of the quantum dimensions of the representations in the su(2)$_k$ anyon theory
\begin{align}
\label{eq:Weight su(2)_k}
W_1(\boldsymbol \alpha,V)= \left\{
\begin{array}{cc}  
2\mathcal{D}^V \prod_{j=0}^k {d_j }^{n_j} & \mbox{if } (\sum_{s=1}^\ell \alpha_s)/2 \in \mathbb{N} \\
0 & \mbox{otherwise}
\end{array}\right. ,
\end{align}
where  $n_j$ is the number representations of type $j$ in the boundary configuration. Deriving Eq.~\eqref{eq:Weight su(2)_k} makes use of the fact that the eigenvalues of the weight matrices are related to the quantum dimensions, which we derive in Appendix \ref{sec:stringnet}. Crossing symmetry implies that the ordering in the boundary configuration is irrelevant, only the number of each type of representation enters the expression for the weight. The summation over all boundary configurations can be performed by noting that the summands are nothing but multinomial coefficients of $\left(\mathcal{D}^{\pm}_{n+1}\right)^\ell$ with 
\begin{align}
\label{eq:D pm}
\mathcal{D}^{\pm}_{m}=\sum_{j=0}^{k}(\pm 1)^j {d_j}^{m}\, 
\end{align}
and the Renyi entropy becomes 
\begin{align}
\label{eq:Renyi single}
S_n^A & =V_A\log \mathcal{D} +\frac{2n\ell}{n-1}\log \mathcal{D} +\frac{1}{1-n}\log\left[\mathcal{D}^{+}_{n+1} \!^{\ell}+ \mathcal{D}^{-}_{n+1}  \!^{\ell}\right]\nonumber\\
 & \approx V_A\log \mathcal{D} +\ell \log\left[\frac{ \mathcal{D}^{2n}}{\mathcal{D}^{+}_{n+1} } \right]^{1/(n-1)}\, ,
\end{align}
where the second line is valid in the limit of a long boundary length $\ell$. 

Note that there is no constant contribution to the Renyi entropy, when part $A$ and $B$ are both connected. The constraint of the boundary --- there are only $\frac 1 2 k^{\ell}$ instead of $k^{\ell}$ boundary configurations that contribute to the entropy --- is exactly compensated by the constraint from the volume, i.e. the factor 2 in the boundary weights $W_1(\boldsymbol \alpha, V)$ in Eq.~\eqref{eq:Weight su(2)_k} as well as  the partition function \eqref{eq:partition function} in Eq.~\eqref{eq:Zsu(2)}.  Note that this additional factor came about because there are two Abelian representations in the su(2)$_k$ model, which resulted in that the highest eigenvalue of  $T$-matrix \eqref{eq:T matrix} was doubly degenerate. 

Let us now assume that part $A$ is connected, but  there are two disconnected regions in $B$ denoted by $B_1 $ and $B_2 $,  e.g. bipartition $A_1$ in Fig.~\ref{fig:LevinWenPartitions}.   The boundary configuration for $B_1(B_2)$ is given by  $\boldsymbol \alpha_1 (\boldsymbol \alpha_2)$ with length $\ell_1(\ell_2)$. The expression for the weights $W_1(\boldsymbol \alpha, V)$ in  Eq.~\eqref{eq:Weight su(2)_k} is unchanged, except that the weight for part $A$ has a boundary that is the combination of $\boldsymbol \alpha_1$ and $\boldsymbol \alpha_2$. That the boundaries combine and can be written  as one boundary of length $\ell=\ell_1+\ell_2$ is ensured by crossing symmetry. The calculation of the Renyi entropy proceeds along the same lines as the one above, but in this case the boundary and volume constraints do not compensate each other. Instead the constraints arising from the volume are only partly compensated by the boundary, which yields an additional $-\log 2$ to the previous result:
\begin{multline}
\label{eq:Renyi double}
S_n^A  =V_A\log \mathcal{D} +\frac{2n\ell}{n-1}\log \mathcal{D} \\
+\frac{1}{1-n}\log\left[\sum_{\sigma_1,\sigma_2=\pm 1} \mathcal{D}^{\sigma_1}_{n+1} \!^{\ell_1}\mathcal{D}^{\sigma_2}_{n+1} \!^{\ell_2}  \right]-\log 2 \\
  \approx V_A\log \mathcal{D} +\ell \log\left[\frac{ \mathcal{D}^{2n}}{\mathcal{D}^{+}_{n+1} } \right]^{1/(n-1)}-\log 2\, ,
\end{multline}
where  the second line is valid when  $\ell_1$ and $\ell_2$ are both large. 
If we had computed $S_n^B$ instead, we had recovered expression \eqref{eq:Renyi single} with $A$ and $ B$ interchanged. 

The calculation above is straightforward to generalize to arbitrary number of regions in $A$ and $B$. We find that in the limit, where  the individual  volumes and boundary lengths are all large, the Renyi entropy is given by:
\begin{align}
\label{eq:final result}
S_n^A & = V_A\log \mathcal{D} +\ell \log\left[\frac{ \mathcal{D}^{2n}}{\mathcal{D}^{+}_{n+1} } \right]^{1/(n-1)} -(n_b-1)\log 2\, ,
\end{align}
where $n_b$ is the number of disconnected regions in part $B$. Analytic continuation to $n=1$, gives the following result for the Shannon entropy:
\begin{multline}
\label{eq:Shannon entropy multiple partition}
S_A =V_A\log \mathcal{D}+\ell\log\left[\mathcal{D}^{2}\prod_{j=0}^{k}(d_{j})^{-d_{j}^{2}/\mathcal{D}^{2}}\right]\\ -(n_b-1)\log 2\, .
\end{multline}

Let us emphasize again that the total quantum dimension $\mathcal{D}$ of the anyon theory appears as coefficient in the volume term, but not in the topological entropy. The latter is instead given by the logarithm of the number of Abelian representations, which is two for {\em all} su(2)$_k$ anyon theories. In Appendix \ref{sec:effective_loop_model} we present an alternative description of the su(2)$_k$ models, which gives an intuitive understanding of this, at first glance surprising, finding that the topological entropy does not depend the level $k$. It turns out that all classical stringnet models based on the su(2)$_k$ anyon theories can be mapped to generalized loop models, for which an topological entropy of $\log 2$ is expected, independent on the details of the loop model.  


\section{Outlook}\label{sec:outlook}

So far, we have focused our discussion on classical stringnets rooted in the su(2)$_k$ anyon theories. However, the quantum double construction of Levin and Wen and as such also the construction of their classical analogs can in principle be applied to a much broader class of anyonic theories (or in more general words topological quantum field theories). 
For the su(2)$_k$ anyon theories we have found that the topological entropy is 
$\gamma_{\rm classical} = \log 2$, 
independent of the level $k$ of the theory.
In fact, this result can readily be generalized to a broader class of anyon theories to
\[
	\gamma_{\rm classical} = \log M \,,
\]
where $M$ is the number of Abelian representations in the underlying anyon theory.
It might thus be interesting to go to anyonic theories that contain more than the strictly two Abelian representations in the su(2)$_k$ theories. One prominent example is the family of su(N)$_k$ anyon theories, which we will touch on in the next section. Another example would be models based on finite groups $G$, which we will briefly discuss in the subsequent section.

\subsection{su(N)$_k$ models}
We first consider stringnet models based on the  su(N)$_k$ anyon theories, which are related the the more conventional SU(N) algebra, in that only a finite number of representations is kept -- similar to the su(2)$_k$ deformation of SU(2) discussed before. The theory contains $\binom{k+N-1}{N-1}$ representations, $N$  of which are Abelian.  
The fusion rules can again be constructed -- similar to the fusion rules \eqref{eq:fusion rules} of su(2)$_k$ -- in a consistent way that incorporates the cutoff $k$ of the deformation, though no closed expression are known for $N>3$.

Constructing the classical stringnet model based on these su(N)$_k$ anyon theories we can again define a set of admissible lattice configurations, where the fusion rules are fulfilled at every single vertex. 
A (weighted) sum over these admissible lattice configurations will then define the partition function of the classical model.
Similar to our discussion of the su(2)$_k$ stringnet models we define the local Boltzmann weight of a vertex via the fusion rules as 
\[
 w(i,j, l)=\mathcal{N}_{ij}^{\bar l} \,,
\]
which not only enforces the fusion rules at the vertex, but also assigns weights to the various admissible vertex configurations. 
In particular, it should be noted that for $k>2$  not all vertices enter with the same weight, but depending on the multiplicity of certain fusion channels $\mathcal{N}_{ij}^l$ the weights can actually vary -- in contrast to the previously discussed case of su(2)$_k$ anyon theories.

The so-defined classical su(N)$_k$ stringnet model obeys the crossing symmetry relation \eqref{eq:crossing symmetry}
\cite{SimonFendley}.  However, as a consequence of the unequal Boltzmann weights for $k>2$ the replicated system used for  calculation of the Renyi entropies no longer obeys the crossing symmetry. As such, our results can be directly generalized only to the case of su(N)$_1$ and su(N)$_2$ anyon theories, while the su(N)$_k$ theories with $k>2$ require additional work beyond the scope of the current manuscript.

Let us first get some intuition by discussing the su(N)$_1$ anyon theory, which turns out to be rather accessible. It has $N$ representations, which can be labeled consecutively by integers $0,\ldots, N-1$. All representation are Abelian with fusion rules 
\begin{align}
\label{eq:su(r)_1}
i\times j= (i+j) \mod N \,,
\end{align} 
where we note that these fusion rules are equivalent to the group operation of the cyclic group $\mathbb Z_N$.

The fusion matrices are then given by 
\begin{align}
\phi^{(\alpha)}_{ij}&=\delta_{(\alpha+i) \bmod N,j} 
\end{align} 
and the $T$-matrix is diagonal with
\begin{align}
\label{eq:T Zk}
T&=N \, \mathbf{1}_N\, .
\end{align}
From the definition of the weight \eqref{eq:Weight} one can deduce that only configurations, where the boundary representations fuse to the identity,  contribute to the Renyi entropies
\begin{align}
\label{eq;Weight Zk}
W_1(\boldsymbol \alpha, V)&= \left\{ \begin{array}{cc} N^{V/2+1}  & \mbox{if }(\sum_{j=1}^\ell \alpha_j)\bmod N =0\\
0 & \mbox{otherwise} \,.
\end{array}\right.
\end{align}
Inserting the weights into  \eqref{eq:Renyi entropy crossing symmetry} and noting that there are $N^{\ell-1}$ allowed boundary configurations for each individual boundary, one obtains the Renyi entropy
\begin{align}
\label{eq:Renyi Zk}
S_n^A&=V_A \log \sqrt N +\ell \log N -(n_b-1)\log N \,,
\end{align}
where  $n_b$ is the number of disconnected regions in $B$. We again see that the Renyi entropy follows a volume law augmented by a boundary term and a topological correction of the form
\begin{align}
	\gamma_{\mbox{\small su(N)}_1} = \log N \,.
\end{align}
This result is precisely in line with the statement that only the number of Abelian anyons in the underlying anyonic theory contributes to the topological correction.

The calculation for su(N)$_2$ is substantially more technical, but yields precisely the same value for the topological entropy
\begin{align}
	\gamma_{\mbox{\small su(N)}_2} = \log N \,.
\end{align}
The interested reader is referred to appendix \ref{sec:stringnet_su(N)} for some of the details of this calculation.

\subsection{Finite groups}
\label{sec:finite groups}
In addition to anyon models described so far, the quantum double construction, the formulation of its classical analog as well as the methods described to calculate the classical entropies in  section~\ref{sec:Renyi entropy} can also be used to study classical stringnet  models based on finite groups, both Abelian and non-Abelian. 
Such a group is  denoted by $G$ in the following and its elements by $e,i,j,\ldots \in G$ with  $e$ being the identity element. The fusion rules are replaced by the usual group operation $\circ$, such that 
\[
   i\circ j=l
\]
implies a `fusion coefficient`  
\[
   \mathcal{N}_{ij}^l=\delta_{i\circ j\circ \bar l,e} \,,
\] 
where $\bar i$ is the (unique) inverse of $i$ defined by $i\circ \bar i=\bar i\circ i=e$. 
The Boltzmann weights are thus defined as 
\begin{align}\label{eq:def weight group}
w(i,j,l)=\left\{
\begin{array}{cc}
1 & \mbox{if } i\circ j\circ l=e\\
0 &\mbox{otherwise}\,.
\end{array} \right.
\end{align}
We want to emphasize, the outcome of `fusing' two particles is always unique given an ordering. Thus, all elements are Abelian according to our previous definition counting the number of fusion outcomes, even though the underlying group may be non-Abelian. In particular, the individual quantum dimensions are all $d_j=1$ and the total quantum dimension $\mathcal{D}=\sqrt{|G|}$, where $|G|$ is the total number of elements in the group.  The calculation of the Renyi entropy is very similar to the one outlined  for su(N)$_1$. In particular,  Eq.~\eqref{eq:Renyi Zk}  still holds when replacing $N$ by $|G|$
\begin{align}
\label{eq:Renyi group}
S_n^A&=V_A \log \sqrt {|G|} +\ell \log |G| -(n_b-1)\log |G| \,.
\end{align}

Finally, we should emphasize that our results for classical models based on finite groups are fully consistent with the results previously reported by Castelnovo and Chamon in Ref.~\cite{CastelnovoChamon1}. See also Ref. \cite{RahmaniChern} for related work on Kagome spin-ice. 

%

\section{Summary}
\label{sec:conclusions}
To summarize our results, the manuscript at hand provides a detailed introduction of the construction of classical stringnet models from quantum double models based on a given anyon theory. For the family of su(2)$_k$ anyon theories, we have carefully analyzed the topological entropies arising as subleading contribution in the Renyi entropies. In particular, we have derived the precise form of the volume law governing the Renyi entropies of order $n$ for a subsystem $A$
\begin{align}
\label{eq:Renyi entropy final}
S_n^A=V_A\log \mathcal{D} +\ell \log\left[\frac{ \mathcal{D}^{2n}}{\mathcal{D}^{+}_{n+1} } \right]^{\frac{1}{(n-1)}}-(n_b-1)\log 2 \, ,
\end{align}
where  $n_b$ is the number of disconnected regions in part $B$, $\mathcal{D}$ is the total quantum dimension \eqref{eq:tot qu dim} of the anyon theory and $\mathcal{D}^+ _{m}=\sum_{j=0}^k d_j ^m$.   

Analytic continuation to $n=1$ yields the Shannon entropy
\begin{multline}
\label{eq:Shannon final}
S_A =V_A\log \mathcal{D}+\ell\log\left[\mathcal{D}^{2}\prod_{j=0}^{k}(d_{j})^{-d_{j}^{2}/\mathcal{D}^{2}}\right]\\ -(n_b-1)\log 2\, .
\end{multline}

These results for the su(2)$_k$ anyon theories can readily be generalized to su(N)$_k$ theories with $k=1,2$ as well as models 
based on finite groups to give a topological entropy of 
\begin{align}
\label{eq:S_top final}
	\gamma = \log M ,
\end{align}
where $M$ is the number of Abelian representations in the underlying anyon theory.  Eq.~\eqref{eq:S_top final} is the main result of our manuscript. 

Let us finally reemphasize  that the classical and quantum variants of topological entropy are encoding substantially different aspects of the system, even though they are defined in a very similar manner. While the quantum topological entropy arises from a constraint on the boundary, the classical entropy rather originates from a constraint on the volume. In particular, in the quantum system the topological $O(1)$ contribution  is proportional to the number of individual boundaries. For the classical counterpart, the Shannon (or Renyi) entropy of subsystem $A$ has a topological contribution, which is proportional to the number of disconnected regions in subsystem $B$, or rather $n_b-1$. Even though the actual values of the classical and quantum topological entropy can turn out to be the same, in particular when considering classical models based on finite groups \cite{CastelnovoChamon1},  they are in general sensitive to different features in the  topological field theory. While the quantum version is sensitive to all representations in the quantum double model of the underlying anyon theory, the classical one is sensitive only to the Abelian ones in the anyon theory-- resulting in vastly different estimates for su(2)$_k$ anyon theories with $k \geq 2$.

Unfortunately, the proof leading to \eqref{eq:S_top final} does not directly generalize to su(N)$_k$ for arbitrary $k$, as the summation over configurations in part $A$ and the boundary cannot be performed analytically. The form of Eq.~\eqref{eq:Renyi entropy crossing symmetry} suggests that the topological contribution is again given by $(n_b-1)\log N$.  However, within our current approach we cannot rigorously proof that there are no other contributions to the topological entropy arising, e.g., from the summation over boundary configurations. For these cases numerical simulations might be useful to shed more light on this question. 

 An important issue that was not discussed in this manuscript is the one of stability. The main feature of the topological entropy of quantum systems is that it is robust against any kind of local perturbations. It was already noted in Ref.~\cite{CastelnovoChamon1} that the classical topological entropy is not robust against softening the vertex constraint. Let us for instance consider the classical loop model. As soon as there is a finite (even if infinitessimal) probability of open loops, the topological entropy vanishes for large enough system sizes. Another important perturbation in the classical stringnet model is introducing a string tension.
Some guidance on this issue arises from the classical loop model. The latter is dual to the 2D Ising model with the loop tension in the loop model corresponding to finite temperature in the Ising model. As such we know that the (topologically non-trivial) loopgas phase persists up to a finite, critical loop tension, corresponding precisely to the critical temperature of the dual Ising model.
As such it is reasonable to expect that the topological entropy remains constant up to the critical value of the loop tension. 
The topological entropy of the classical system can then be used to characterize an entire phase in full analogy to its quantum counterpart. \\

\noindent{\bf Acknowledgements --}
M.H. thanks Eddy Ardonne, Steve Simon, and Joost Slingerland for interesting and stimulating discussions. 
S.T. thanks R.G. Melko for an inspiring discussion that has led to the idea for the calculations in the manuscript at hand. We also  thank T. Quella for a critical reading of the manuscript. 
We acknowledge partial support from SFB TR 12 of the DFG. 

%

\appendix

%

\section{Renyi entropy for the su(2)$_k$ stringnet}
\label{sec:stringnet}
In this appendix we derive the analytic expression of the Renyi entropies for su(2)$_k$ stringnets in the limit of large volume. We focus on a bipartition, where part $A$ and $B$ are both connected. Generalizing to disconnected regions in $B$ and/or $A$ is straightforward. 

The weight matrices and the $T$ matrix are simultaneously diagonalized by the modular $S$-matrix, which is known explicitly for su(2)$_k$, see e.g. Ref.~\cite{yellow_book}: 
\begin{align}
\label{eq:s matrix}
S_{ij} & =\sqrt{\frac{2}{k+2}}\sin\left(\frac{(i+1)(j+1)}{k+2}\pi\right)
\end{align}
with $i,j=0,\ldots,k$. 
The explicit form of the $S$-matrix is in fact not important for the proof.  
However, we will use that the $S$-matrix elements are directly related to the quantum dimensions $d_j$ of the representations of the su(2)$_k$ anyon theory: 
\begin{align}
\label{eq:S matrix props}
S_{00}&=\mathcal{D}^{-1}\nonumber\\
S_{j0}&=S_{0j}=\frac{d_j}{\mathcal{D}}\,, 
\end{align}
where $\mathcal{D}$  is the total quantum dimension \eqref{eq:tot qu dim}.
Consequently, the eigenvalues of the weight matrices and the $T$-matrix 
\begin{align}
\label{eq:evs}
\left[S^\dagger \phi^{(\alpha)} S\right]_{i,j}&=\delta_{i,j}\lambda_{j}^{(\alpha)}  \nonumber\\
\left[S^\dagger T S\right] _{i,j}&=\delta_{i,j} t_{j}  
\end{align}
are also related to the quantum dimensions: 
\begin{align}
\label{eq:ev_quantumdimensions}
\lambda_{j}^{(\alpha)}=\frac{S_{\alpha j}}{S_{0j}}\nonumber\\
 t_{j}  =\left(\frac{\mathcal{D}}{d_{j}}\right)^{2}\,.
\end{align}
In particular, 
\begin{align}
\label{eq:ev_qd}
\lambda^{(\alpha)}_0&=d_\alpha\nonumber\\
\lambda^{(\alpha)}_k&=(-1)^\alpha d_\alpha\, ,
\end{align}
 which can readily be derived from \eqref{eq:s matrix}. 
For su(2)$_k$ there are exactly two Abelian representations, labeled by $0$ and $k$, with $d_{0}=d_{k}=1$. 
All other represenations are non-Abelian and have, therefore, larger quantum dimensions. 
Thus, the highest eigenvalue of the $T$-matrix has value $\mathcal{D}^2$ and is two-fold degenerate. 
 As a consequence, we find that the partition function \eqref{eq:partition function} grows asymptotically as a power of the total quantum dimension $\mathcal{D}$ 
 \begin{align}
 Z[V]=\sum_{j=0}^k \left(\frac{\mathcal{D}}{d_j}\right)^V\approx 2\mathcal{D}^V
 \end{align}
in the limit of large $V$. 

Let us now proceed by deriving an explicit formula for the boundary weight \eqref{eq:Weight}. In the following, we assume that the volumes of both subsystems are large. The boundary weight of a boundary configuration $\boldsymbol\alpha$ and volume $V$ is by definition 
\begin{align}
W_1(\boldsymbol\alpha,V)&=\sum_{\beta=0}^k \left[\phi^{(\alpha_1)} \ldots \phi^{(\alpha_\ell)}\right]_{0,\beta} \mbox{Tr}\left[\phi^{(\beta)} T^{V/2}\right]\nonumber\\
&=\sum_{\beta,s,j=0}^k S_{0j} \lambda^{(\alpha_1)} _j \ldots \lambda^{(\alpha_\ell)}_j S_{\beta j}  \lambda^{(\beta)}_s\left(\frac {\mathcal{D}} {d_s}\right)^V \, .
\end{align}
Using the explicit expression of $\lambda^{(\beta)}_s$ in terms of the $S$-matrix elements, performing the summation over $\beta$ and noting that $SS^\dagger =S^\dagger S=\mathbf{1}_{k+1}$ yields 
\begin{align}
W_1(\boldsymbol\alpha,V)&=\sum_{s,j=0}^k \delta_{j,s}  \lambda^{(\alpha_1)} _j \ldots \lambda^{(\alpha_\ell)}_j   \left(\frac {\mathcal{D}} {d_s}\right)^V\, . 
\end{align}
In  the large volume limit only the terms with $s=0,k$ survive, the others are exponentially suppressed. By noting that  Eq.~\eqref{eq:ev_qd}  relates the $0$th and $k$th eigenvalue of the weight matrices to the quantum dimension, we arrive at the final result for the boundary weight
\begin{align}
\label{eq:Weight_su(2)k}
W_1(\boldsymbol \alpha,V) 
 & =(1+(-1)^{\Sigma})\mathcal{D}^{V}\prod_{j=1}^{\ell}\lambda_{0}^{(\alpha_{j})}\nonumber\\
&=(1+(-1)^{\Sigma})\mathcal{D}^{V}\prod_{j=0}^{k}{d_j}^{n_{j}} \, ,
\end{align}
where $\Sigma=\sum_{j=0}^{\ell}\alpha_{j}$ and $n_{j}$ is the multiplicity of the representation labeled by $j$ on the boundary. The ordering of the representations is unimportant due to crossing symmetry. 

Using the explicit form of the boundary weights, the Renyi entropy can be computed straightforwardly.   
\begin{align}
\label{eq:Renyi entropy single partition1}
S_{n}^A & =\frac{1}{1-n}\log\left[\sum_{\boldsymbol \alpha}W_1(\boldsymbol \alpha,V_A)\left(\frac{W_1(\boldsymbol \alpha,V_B)}{Z}\right)^{n}\right] \nonumber\\
 & =\frac{1}{1-n}\log\left[\sum_{n_{0}+\ldots+n_{k}=\ell}\frac{\ell!}{n_{0}!\ldots n_{k}!}\right. \nonumber\\
 & \left. \times \prod_{j=0}^{k}\left(d_j\right)^{n_{j}(n+1)} (1+(-1)^{\Sigma})^{n+1}\right]\nonumber\\
 &+\frac{1}{1-n}\log\left[\mathcal{D}^{V_A+nV_B-n(V_A+V_B+2\ell)}2^{-n}\right] \,,
\end{align}
where the binomial factors enumerate the number of configurations for given $n_0,\ldots ,n_k$.  Using $(1+(-1)^\Sigma)^{n+1}=2^n (1+(-1)^\Sigma)$ and $\Sigma=\sum_{j=0}^k jn_j$ we find that 
\begin{align}
\label{eq:Renyi entropy single partition2}
S_{n}^A  & =V_A\log \mathcal{D}+\frac{2n\ell}{n-1}\log \mathcal{D} \nonumber\\
 & +\frac{1}{1-n}\log\left[\sum_{n_{0}+\ldots+n_{k}=\ell}\frac{\ell!}{n_{0}!\ldots n_{k}!}\prod_{j=0}^{k}\left(d_{j}^{(n+1)}\right)^{n_{j}}\right. \nonumber\\
  & \left. + \sum_{n_{0}+\ldots+n_{k}=\ell}\frac{\ell!}{n_{0}!\ldots n_{k}!}\prod_{j=0}^{k}\left((-1)^j d_{j}^{(n+1)}\right)^{n_{j}}\right]\, ,
\end{align}
 where the last two lines are multinomial expansions of $\left(\mathcal{D}_{n+1}^\pm\right)^\ell$ with 
\begin{align}
\label{eq:DpmApp}
\mathcal{D}^{\pm}_{m}=\sum_{j=0}^{k}(\pm 1)^j (d_j )^{m}\, .
\end{align}
We can simplify Eq.~\eqref{eq:Renyi entropy single partition2} by noting that the contribution from $\mathcal{D}_{n+1}^-$ vanishes exponentially in the limit of large $\ell$ and obtain
\begin{align}
\label{eq:Renyi entropy single partition large boundary}
S_{n}^A & =V_A\log \mathcal{D}+\ell\log\left[\frac{\mathcal{D}^{2n}}{\mathcal{D}^{+}_{n+1} }\right]^{1/(n-1)}\, .
\end{align}
 In the limit $n\rightarrow1$, we find the following limiting behavior for the Shannon entropy:
\begin{align}
\label{eq:Shannon entropy single partition}
S_{n\rightarrow1} & =V_A\log \mathcal{D}+\ell\log\left[\mathcal{D}^{2}\prod_{j=0}^{k}(d_{j})^{-d_{j}^{2}/\mathcal{D}^{2}}\right]\, .
\end{align}

%

\section{Renyi entropy for the su(N)$_2$ stringnet}
\label{sec:stringnet_su(N)}
Let us now briefly comment on the su(N)$_k$ models and give the final result for the Renyi entropies of the stringnet models based on the su(N)$_2$ anyonic theories. The calculation proceeds along the same lines as in the previous appendix. Thus, we only comment on the few details that differ from the previous calculation. 

The su(N)$_k$ anyonic theories have $N$ Abelian representations. A similar calculation as the one done in Eqs.~\eqref{eq:evs} and \eqref{eq:ev_quantumdimensions} shows that the $T$-matrix has an $N$-fold degenerate, highest eigenvalue $\mathcal{D}^2$. In the su(2)$_k$ models, we saw that the eigenvalues of the weight matrices have special properties, namely $\lambda^{(\beta)}_0=(-1)^\beta \lambda^{(\beta)}_k=d_\beta$, where 0 and $k$ were labeling the Abelian representations. A similar property is valid for su(N)$_k$. In order to see this let us first introduce some necessary notation. In the  su(N)$_k$ anyon theories the representations can be labeled by  $N$ component vectors $\boldsymbol \alpha=(\alpha_{0},\ldots ,\alpha_{N-1})$ \cite{yellow_book}, such that the sum of all the vector entries is $k$.  The Abelian representations are labeled by the vectors, where one component is $k$ and all the others are 0. For instance, we can label all the Abelian representations consecutively by  vectors $\boldsymbol \mu_j=(\mu_{j0},\ldots,\mu_{j N-1})$ with $j=0,\ldots N-1$ and  $\mu_{j i}=k\delta_{j i}$, where  $\boldsymbol\mu_0=(k,0,\ldots, 0)$ denotes the identity representation. It can be shown by using the outer automorphism of su(N)$_k$ that some of the eigenvalues of the weight  matrices are related by phases \cite{yellow_book}:
\begin{align}
\label{eq:S phases}
\lambda_{\boldsymbol\mu_j}^{(\boldsymbol\alpha)} = \exp\left[2\pi i \sum_{s=0}^{N-1} s\alpha_s  \frac j N\right] \lambda_{\boldsymbol\mu_0}^{(\boldsymbol\alpha)}\, . 
\end{align}
Note that in contrast to the index $\boldsymbol\mu_j$, which labels an Abelian representation, $\boldsymbol\alpha$ may label any representation, in particular it may label one of the non-Abelian ones. The phases are $N$th roots of unity in analogy to what was found for the  su(2)$_k$ anyon theory. In analogy to the previous calculation, we can identify $\lambda^{(\boldsymbol \alpha)}_{\boldsymbol\mu_0}$ with the quantum dimension of the representation labeled by $\boldsymbol \alpha$, i.e.  $\lambda^{(\boldsymbol \alpha)}_{\boldsymbol\mu_0}=d_{\boldsymbol \alpha}$.

The boundary weights are computed in same way as shown in Appendix \ref{sec:stringnet}. Restricting the discussion to systems where the volumes of each of the subsystem is large and using \eqref{eq:S phases} to relate the eigenvalues of the weight matrices to the quantum dimensions, we find the following form of the boundary weights 
\begin{align}
W_1(\boldsymbol \alpha,V) 
&\approx \sum_{n=0}^{N-1}\exp\left[\frac {2\pi in}{N}\sum_{j=1}^{\ell}\sum_{s=0}^{N-1} s \alpha_{js} \right] \prod_{i=1}^{\ell}{d_{\boldsymbol \alpha_i}} \, \mathcal{D}^{V}\nonumber\\
&=\left \{
\begin{array}{cc}
N  \prod_{i=1}^{\ell}{d_{\boldsymbol\alpha_i}} \, \mathcal{D}^{V} & \mbox{ for } \sum_{j=1}^\ell \sum_{s=0}^{N-1} \frac{s \alpha_{js}}{N} \in \mathbb{N} \\
 0 & \mbox{ otherwise }\, . 
 \end{array}
\right.  
\end{align}
We note that the form of the boundary weight is very similar to what was found in Appendix \ref{sec:stringnet}. 
The remaining calculation proceeds analogously to the one in Appendix~\ref{sec:stringnet} and the scaling form of the Renyi entropy -- in the limit of large volumes and boundaries -- looks very similar to the result of the stringnet based on the  su(2)$_k$ anyon theory
\begin{align}
\label{eq:Renyi_su(N)k}
S_{n}^A & =V_A\log \mathcal{D}+\ell\log\left[\frac{\mathcal{D}^{2n}}{\mathcal{D}^{+}_{n+1} }\right]^{1/(n-1)}\!\!-(n_b-1)\log N\, ,
\end{align}
except that the total quantum dimension is now the one of the su(N)$_2$ anyon theory and the topological contribution is proportional to $\log N$, as there are  $N$ Abelian representations in su(N)$_2$. The number of disconnected  regions in subsystem $B$ is again denoted by $n_b$. $\mathcal{D}^\pm_m$ is defined as in Eq.~\eqref{eq:DpmApp}.

\section{Effective loop model}
\label{sec:effective_loop_model}
\begin{figure}[t]
  \includegraphics[width=\columnwidth]{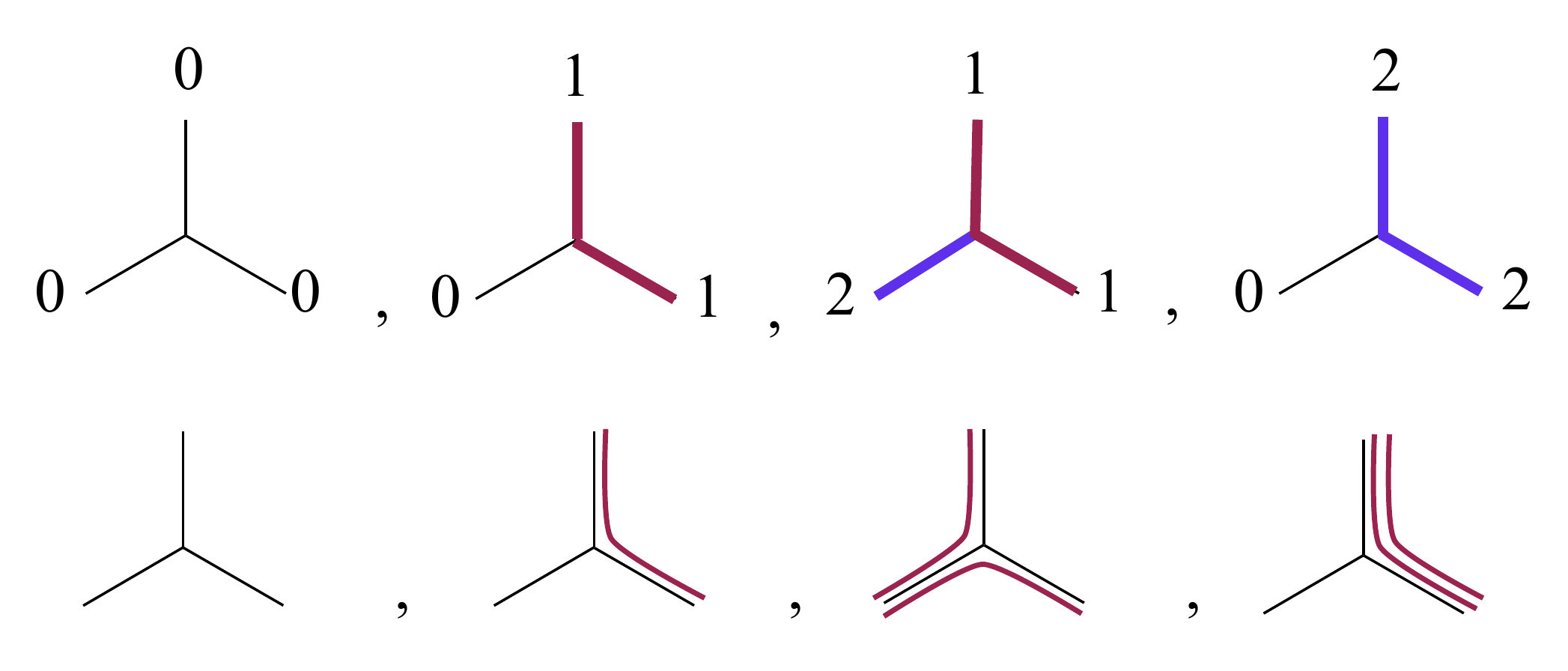}
  \caption{(color online) Allowed vertices up to rotation for su(2)$_2$ stringnets using a generalized loop model. 
  	        The vertices are in one-to-one correspondence to the ones in panel b) of Fig.~\ref{fig:allowed_vertices}.
	        Note that a vertex with two lines on each outgoing leg is absent in the theory. }
  \label{fig:su(2)_2loop}
\end{figure}
In this appendix, we want to present an alternative description on the classical stringnets based on the su(2)$_k$ anyon theories that is useful for understanding why the classical topological entropy is independent on the level $k$. The main idea is to map the   su(2)$_k$ stringnets  to  models of non-crossing loops, where at most $k$ loops are allowed on each edge. The correspondence between the classical stringnet and this generalized loop model is illustrated in Fig.~\ref{fig:su(2)_2loop} for the case of $k=2$ --- the upper panel shows the allowed vertices for su(2)$_2$ and the lower panel the corresponding vertices in the loop model. In order to reproduce the  allowed vertices for su(2)$_k$ one needs to put additional constraints on the non-crossing loop model. For instance, the loop configuration with two strings on each edge is not allowed.

The topological entropy is by definition a quantity that is not sensitive to local details. In contrast, it indicates a global conservation law, which on the classical level is enforced by local (hard) constraints. For loop models, this constraint is the absence of open strings. Thus, as long as the partition function contains loops on all length scales with a finite weight, the topological entropy should be given by $\log 2$ -- independent of which particular loop model is studied. Thus, it seems reasonable that the topological entropy of all su(2)$_k$ models is given by  $\gamma=\log 2$.


\begin{thebibliography}{99}
\bibitem{FootnoteNonAbelianess} 
In even more general terms, a representation is called non-Abelian if fused with any other representation of the same anyonic theory leads to more than one fusion outcome.
\bibitem{FootnoteToricCode}
Note the subtle difference to the loopgas configurations of the toric code model, which are equal-weight superpositions of closed loop configurations, similar to those occurring for the su(2)$_1$ anyon theory, where all coefficients $a_s$ are strictly positive and of identical magnitude.
\end{thebibliography}
\end{document}